\documentclass[aps,prd,onecolumn,nofootinbib,superscriptaddress]{revtex4}
\pdfoutput=1
\usepackage{graphicx}
\usepackage{amsmath}
\usepackage{amsfonts}
\usepackage{amssymb,ulem}
\usepackage{color}%
\usepackage{dcolumn}

\usepackage{MnSymbol,wasysym}
\usepackage{braket}
\usepackage{eurosym}
\usepackage{calrsfs}
\usepackage{multirow}
\usepackage{float}
\usepackage[usenames,dvipsnames,svgnames]{xcolor}

\newcommand{\RNum}[1]{\uppercase\expandafter{\romannumeral #1\relax}}
\usepackage[colorlinks=true,linkcolor=blue,urlcolor=blue,filecolor=black,citecolor=red,pdfstartview=FitV,pdftitle={},pdfsubject={},pdfkeywords={},pdfpagemode=None,bookmarksopen=true]{hyperref}
\usepackage[title]{appendix}
\makeatletter
\newcommand*{\rom}[1]{\expandafter\@slowromancap\romannumeral #1@}
\makeatother

\begin{document}
\baselineskip=0.5 cm

\title{Perturbations, eikonal limit and Hawking radiation of diverse external fields in a non-minimal charged black hole}

\author{Yinan Jia}
\email{jiayinan$_$yzu@163.com}
\affiliation{Center for Gravitation and Cosmology, College of Physical Science and Technology, Yangzhou University, Yangzhou, 225009, China}

\author{Zhen-Hao Yang}
\email{yangzhenhao$_$yzu@163.com}
\affiliation{Center for Gravitation and Cosmology, College of Physical Science and Technology, Yangzhou University, Yangzhou, 225009, China}

\author{Xiao-Mei Kuang}
\email{xmeikuang@yzu.edu.cn (corresponding author)}
\affiliation{Center for Gravitation and Cosmology, College of Physical Science and Technology, Yangzhou University, Yangzhou, 225009, China}

\begin{abstract}
\baselineskip=0.4 cm
We compare the quasinormal modes (QNMs), greybody factors, and Hawking emission of massless scalar, electromagnetic, and Dirac test fields propagating on a non-minimal Einstein--Yang--Mills black hole. The QNMs are calculated using WKB and pseudospectral methods, while greybody factors are estimated with sixth-order WKB method. All computed modes are damped within the parameter range considered. Higher overtones generally exhibit stronger dependence on the non-minimal coupling $\xi$ than the fundamental mode, including non-monotonic behavior in the scalar and Dirac sectors. For large multipole numbers, all three fields follow a common leading eikonal description governed by the unstable circular null orbit. Increasing $\xi$ raises the orbital angular velocity and reduces the Lyapunov exponent, leading to higher oscillation frequencies and weaker damping rate. The greybody factors depend strongly on the multipole number, with weaker field dependent corrections from $\xi$. The  partial-wave emission powers generally decrease with increasing $\xi$, while the lowest scalar mode shows mild non-monotonic behavior. This comparison reveals enhanced overtone sensitivity alongside comparatively small changes in transmission, whereas the Hawking emission reflects changes in both the greybody factors and the black hole temperature.
\end{abstract}

\maketitle

\tableofcontents
\section{Introduction}
The detection of gravitational waves (GWs) from binary black hole mergers has opened new opportunities to test gravity in the strong field regime \cite{LIGOScientific:2016aoc,LIGOScientific:2016lio}. During the ringdown stage, the response of the remnant black hole is dominated by damped oscillations known as quasinormal modes (QNMs). Their complex frequencies are determined by the background geometry, the perturbation equations, and the corresponding boundary conditions \cite{Kokkotas:1999bd,Berti:2009kk,Konoplya:2011qqh}. The real part of a quasinormal frequency describes the oscillation frequency, while the negative imaginary part describes the decay of the perturbation. Comparing these frequencies and damping times with theoretical predictions forms the basis of black hole spectroscopy, which provides a means of testing general relativity (GR) and constraining possible deviations from its predictions \cite{Dreyer:2003bv,Berti:2005ys}.

It is commonly known that there remain strong motivations to explore extensions of GR, which often introduce additional fields or higher curvature terms into the action \cite{Nojiri:2006ri,Clifton:2011jh,Berti:2015itd}. These theories provide a broader framework for examining the foundations and limits of GR and investigating gravitational phenomena on astrophysical and cosmological scales. Describing black hole GW ringdown in such theories generally requires an analysis of coupled gravitational and matter perturbations, which is usually technically demanding. Test field perturbations provide a tractable setup for studying how wave propagation responds to changes in the background geometry. Although their spectra do not directly describe GW ringdown, they offer useful information about stability in the corresponding test field sectors and the dependence of wave dynamics on the black hole parameters. They also allow fields with different spins to be compared on the same background.
In addition, the same radial effective potentials determine the transmission probabilities, or greybody factors, that modify the Hawking spectrum measured at infinity \cite{Hawking:1975vcx,Page:1976df}. QNMs and greybody factors therefore provide complementary information about wave propagation in a given black hole spacetime. Moreover, their starting points are similar, only with different boundary conditions to solve the master equation of motions:  QNM requires a purely outgoing wave at infinity and a purely ingoing wave at the event horizon,  while Hawking radiation spectrum allows ingoing waves at infinity.

Among extensions of GR involving non-Abelian matter, Einstein-Yang-Mills (EYM) theory provides an important framework for studying the interaction between gravitational and gauge fields.  An important motivation is that, in four spacetime dimensions, neither asymptotically flat vacuum gravity nor pure Yang-Mills (YM) theory on Minkowski spacetime admits nontrivial, static, globally regular solutions with finite energy. Such solitonic solutions do, however, exist in the coupled EYM system \cite{Bartnik:1988am}. EYM theory also admits black holes with non-Abelian gauge hair that is not fully characterized by the asymptotic conserved charges, providing counterexamples to the generalized black hole no-hair conjecture \cite{Kleihaus:1998sm}. Non-minimal coupling between the gravitational and gauge fields further extends this framework. In particular, exact regular black hole solutions have been constructed in non-minimally coupled EYM theory using an $SU(2)$ Wu--Yang ansatz \cite{Balakin:2015gpq}, which will be reviewed soon. Its
metric depends on the mass $M$, magnetic charge $Q$ and non-minimal coupling $\xi$ . The solution is reduces to the Reissner-Nordström (RN) metric when $\xi= 0$ .
For appropriate parameter choices, the curvature invariants of these solutions remain finite throughout spacetime, making them useful models for investigating black holes without curvature singularities \cite{Lan:2023cvz}. And,  it provides a useful background for examining how a non-minimal interaction affects the strong field physics and test field propagation.

The geodesic and optical properties of this spacetime have been examined in several studies. Particle motion, circular orbits, and black hole images were investigated in Refs.~\cite{Al-Badawi:2023jtq,Chen:2025ujl}, including the dependence of the photon sphere and the innermost stable circular orbit on the model parameters. Weak and strong gravitational lensing were studied in Ref.~\cite{Liu:2019pov}. A rotating metric obtained from the static solution through a modified Newman--Janis procedure has also been used to investigate black hole shadows, QNMs, and quasiperiodic oscillations \cite{Jusufi:2020odz,Ali:2026pmh}. Further studies considered the shadow and weak gravitational lensing in a plasma environment \cite{Kala:2022uog}, as well as strong gravitational lensing by the rotating geometry \cite{Zhang:2023oui}.

The perturbative and radiative properties of the non-minimal EYM black holes have also attracted attention. Scalar filed QNMs and their relation to the black hole shadow were investigated in Ref.~\cite{Gogoi:2024qnm}, while long-lived modes and quasi-resonances of massive scalar fields were studied in Ref.~\cite{Dubinsky:2025bvf}. Greybody factor bounds for scalar and fermionic fields were obtained in Ref.~\cite{Al-Badawi:2023qnm}. An outburst of scalar QNM overtones was subsequently reported in Ref.~\cite{Lutfuoglu:2025ljm}, in which the authors also examnined  scalar greybody factors. These results provide a basis for examining how the field dependent QNM response is related to the scattering and radiative properties.

Motivated by these considerations, we shall investigate massless scalar, electromagnetic (EM), and Dirac perturbations of the non-minimal EYM black hole. All three fields are treated as minimally coupled test fields, with the electromagnetic field taken to be an independent Abelian probe. The non-minimal coupling parameter therefore affects their propagation. We focus on how the sensitivity to $\xi$ changes with the field spin and overtone number, and how the differences at low multipoles connect to the common leading order geometric-optics description. We first derive the master equations and examine the corresponding effective potentials. We then calculate the QNMs using WKB method  \cite{PhysRevD.35.3621,Konoplya:2003ii,Matyjasek:2017psv} and pseudospectral method~\cite{Zenginoglu:2007jw,PanossoMacedo:2023qzp,Yang:2025hqk}, and study the dependence of the fundamental mode and higher overtones on $\xi$ and the multipole number. The large multipole results are compared with the eikonal prediction determined by the angular velocity and Lyapunov exponent of the unstable circular null orbit. Finally, we estimate the greybody factors using the sixth-order WKB approximation and evaluate the corresponding Hawking emission spectra and partial-wave powers. This comparison allows us to examine how the same geometric deformation affects oscillation, damping, and transmission in the three test field sectors.

The remainder of this paper is organized as follows. In section \ref{sec:eoms}, we review the non-minimal EYM black hole and derive the perturbation equations for different test fields. Moreover, we analyze the effective potentials and discuss their dependence on the model parameters.  In section \ref{sec:QNM}, we calculate the quasinormal modes and investigate the effects of the coupling parameter and multipole number. In section \ref{sec:Hawking Radiation},  by solving the corresponding master equations, we evaluate the greybody factor and the energy emission of Hawking radiation for various particles. Finally, we conclude and discuss our results in section \ref{sec:conclusion}. Throughout the manuscript, we consider $c=\hbar=8\pi G=1$.

\section{Master equations and effective potentials of the test fields}\label{sec:eoms}

In this section, we briefly review the background of the non-minimal EYM black hole and derive the master equations for various massless test fields, including scalar, electromagnetic, and Dirac fields.  The fields are minimally coupled to the metric, meaning that their backreaction to the background is neglected. The corresponding effective potentials for different perturbation sectors are obtained, and the effects of  $\xi$ as well as  the multipole number $\ell$  are analyzed.

\subsection{Black hole background}
The action  of the non-minimal EYM theory  is given by \cite{Balakin:2006qe,Balakin:2015gpq}
\begin{eqnarray}\label{eq:action}
 S=\frac{1}{8\pi}\int d^4x\sqrt{-g}\left[R+\frac{1}{2}F_{\mu\nu}^{(a)}F^{\mu\nu(a)}+\frac{1}{2}\bar{R}^{\alpha\beta\mu\nu}
F_{\alpha\beta}^{(a)}F_{\mu\nu}^{(a)}\right],
\end{eqnarray}
where $R$ denotes the Ricci scalar curvature. The Greek indices $\mu$, $\nu$ run from $0$ to $3$, corresponding to the four-dimensional spacetime coordinates, while the Latin indices $a$, $b$, $c$ range from $1$ to $3$ and represent the internal indices of the $SU(2)$ YM gauge group. The quantity $F_{\mu\nu}^{(a)}$ represents the YM field strength tensor, which is related to the gauge potential $A_{\mu}^{(a)}$ through
\begin{eqnarray}
  F_{\mu\nu}^{(a)}=\nabla_\mu A_\nu^{(a)}-\nabla_\nu A_\mu^{(a)}+f_{(b)(c)}^{(a)}
A_\mu^{(b)}A_\nu^{(c)},
\end{eqnarray}
where $\nabla_\mu$ denotes the covariant derivative, and $f_{(b)(c)}^{(a)}$ are the structure constants of the three-parameter YM gauge group. The tensor $\bar{R}^{\alpha\beta\mu\nu}$, which describes the non-minimal coupling between the YM field and the gravitational field, is defined as \cite{Balakin:2015gpq}
\begin{eqnarray}
    \bar{R}^{\alpha\beta\mu\nu}=\frac{\xi_1}{2}R\left(g^{\alpha\mu}g^{\beta\nu}
-g^{\alpha\nu}g^{\beta\mu}\right)+\frac{\xi_2}{2}\left(R^{\alpha\mu}g^{\beta\nu}
-R^{\alpha\nu}g^{\beta\mu}+R^{\beta\nu}g^{\alpha\mu}-R^{\beta\mu}g^{\alpha\nu}
\right)+\xi_3R^{\alpha\beta\mu\nu},
\end{eqnarray}
where $R^{\mu\nu}$ and $R^{\alpha\beta\mu\nu}$ denote the Ricci tensor and the Riemann curvature tensor, respectively. The parameters $\xi_i(i=1,2,3)$ characterize the non-minimal coupling strengths between the YM field and the gravitational field. By adopting the Wu-Yang ansatz for the gauge field and choosing the coupling parameters as $\xi_1=\xi$, $\xi_2=4\xi$ and $\xi_3=-6\xi$ with $\xi>0$, a regular, static, and spherically symmetric black hole solution can be obtained, of which the metric is \cite{Balakin:2006qe,Balakin:2015gpq}
\begin{eqnarray}\label{eq:metric}
ds^2=-f(r)dt^2+\frac{dr^2}{f(r)}+r^2(d\theta^2+\sin^2\theta d\varphi^2)
~~~\mathrm{with}~~~f(r)=1+\left(\frac{r^4}{r^4+2\xi{Q^2}}\right)\left(\frac{Q^2}{r^2}-\frac{2M}{r}\right).
\end{eqnarray}
Here, $M$ denotes the black hole mass, $Q$ is the YM magnetic charge and $\xi$ controls the strength of the non-minimal interaction. When $\xi=0$, the solution reduces to the minimally coupled EYM black hole, while $Q\to 0$, recovers the Schwarzschild solution. Compared with the minimally coupled case, the non-minimal coupling parameter modifies the near-horizon structure of the spacetime, making this black hole solution an interesting background for studying the effects of modified gravity on perturbation dynamics.

Before studying the perturbation dynamics, it is necessary to determine the parameter region in which the spacetime indeed possesses an event horizon. The horizons can be determined by the equation
\begin{equation}
P(r)=r^4-2M r^3+Q^2r^2+2\xi Q^2=0.
\end{equation}
The extremal configuration satisfies  $P(r_e)=0$ and $P'(r_e)=0$, where the inner and outer horizons coincide.  This gives
\begin{eqnarray}
    r_e=\frac{1}{4}(3M+\sqrt{9M^2-8Q^2}),~~~\xi_e=\frac{27M^4-36M^2Q^2+8Q^4+M(9M^2-8Q^2)^{3/2}}{64Q^2}.
\end{eqnarray}
The parameter region admitting black-hole solutions is $0\leq \xi\leq \xi_e$, with equality at RN black hole and extremal limit, respectively. Note that the Schwarzschild limit $Q=0$ should be  treated separately because the expression for $\xi_c$ is singular there.

\begin{figure*}[htbp]
\centering
\includegraphics[height=5cm]{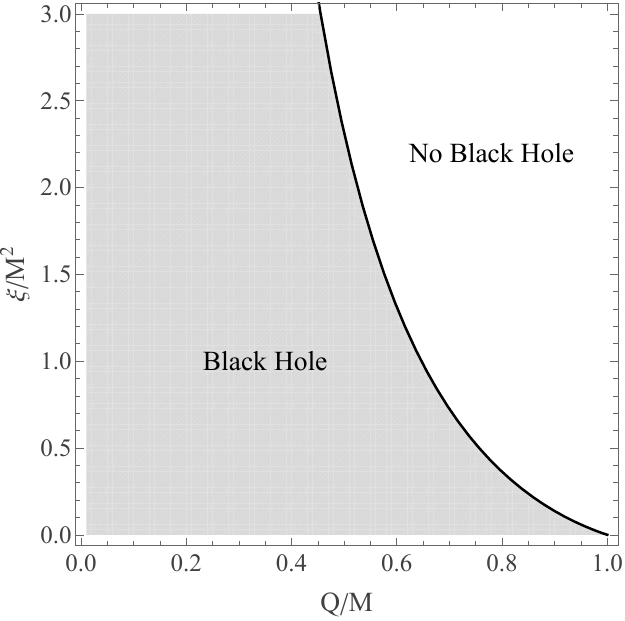}
\caption{\label{Fig:Q-Xi-parameter-space}Allowed parameter region of the non-minimal EYM black hole in the ($Q/M, \xi/M^2$) plane. The shaded area admits an event horizon,  and the solid curve indicates the extremal limit at which the inner and outer horizons merge. The unshaded area has no event horizon. }
\end{figure*}

In FIG. \ref{Fig:Q-Xi-parameter-space}, we show the allowed parameter region of the non-minimal EYM black hole in the ($Q/M, \xi/M^2$) plane. As $Q/M$ increases, the critical value $\xi_c/M^2$ decreases monotonically. This indicates that a larger YM magnetic charge imposes a stronger upper bound on the non-minimal coupling parameter. Conversely, relatively large values of $Q/M\to1$, the critical coupling approaches zero, and the allowed black hole parameter region becomes progressively narrower.
For convenience,  $r$, $\xi$ and $Q$ in the following studies shall be rescaled in the units of $M$, $M^2$ and $M$, respectively. We will fix $M=1$ and $Q/M=1/2$ for which the critical coupling is $\xi_c/M \simeq 2.314$. All values used below lie below this bound, ensuring the existence of an event horizon for the background spacetime. 

\subsection{Scalar field perturbation}
We first consider  a massless scalar field  $\Phi$ minimally coupled to the above background geometry. Its equation of motion is the Klein-Gordon equation
\begin{equation}\label{K--G Eq.}
\square\Phi=\frac{1}{\sqrt{-g}}\partial_\mu(g^{\mu\nu}\sqrt{-g}\partial_\nu\Phi)=0.
\end{equation}
The spherically symmetric metric \eqref{eq:metric} allows the separation
\begin{equation}
\Phi(t,r,\theta,\varphi)=e^{-i\omega t}\frac{R(r)}{r}Y_{\ell m}(\theta,\varphi),
\end{equation}
where $\omega$  is the mode frequency and $Y_{\ell m}$ are the scalar spherical harmonics with the multipole number $\ell=0,1,2,\cdots$ and azimuthal quantum number  $m=-\ell,\cdots,\ell$. The frequency is complex for a QNM and real for a scattering solution.  Substituting it into  \eqref{K--G Eq.}, we obtain the radial master equation
\begin{equation}\label{scalar radial Eq. in r}
f^2(r)\frac{d^2R}{dr^2}+f(r)f'(r)\frac{dR}{dr}+[\omega^2-V_{sc}(r)]R=0,
\end{equation}
with the effective potential
\begin{equation}\label{eq:Vsc}
V_{sc}(r)=\left[1+\frac{r^4}{r^4+2\xi Q^2}\left(\frac{Q^2}{r^2}-\frac{2M}{r}\right)\right]\left[\frac{\ell(\ell+1)}{r^2}+\frac{2(Mr^5-Q^2r^4-6\xi MQ^2r+2\xi Q^4)}{(r^4+2\xi Q^2)^2}\right].
\end{equation}
To simplify the analysis, we introduce the tortoise coordinate
\begin{equation}\label{eq:rstar}
r_*=\int\frac{dr}{f(r)}.
\end{equation}
Then the radial equation takes the Schrödinger--like form
\begin{equation}\label{eq:scalar master eq}
\frac{d^2R}{dr_*^2}+[\omega^2-V_{sc}(r)]R=0.
\end{equation}

\begin{figure*}[htbp]
\centering
\includegraphics[height=4cm]{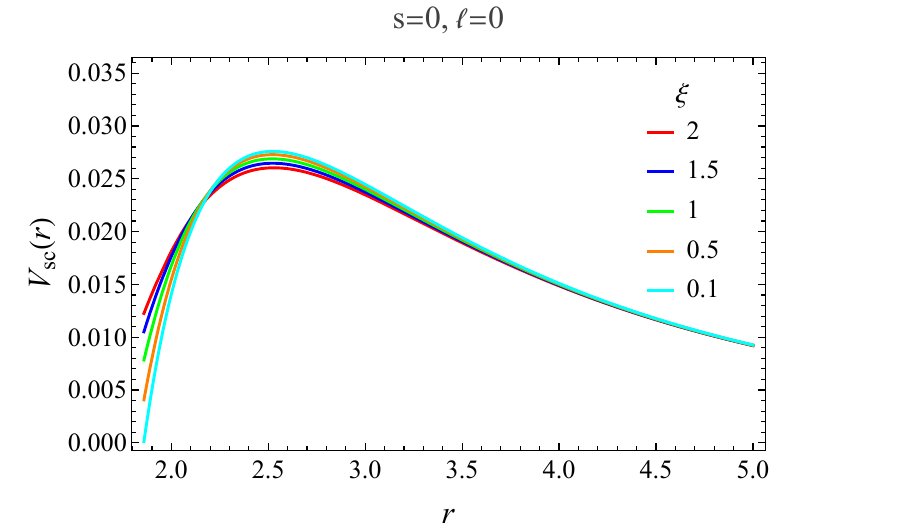}\hspace{0.5cm}
\includegraphics[height=4cm]{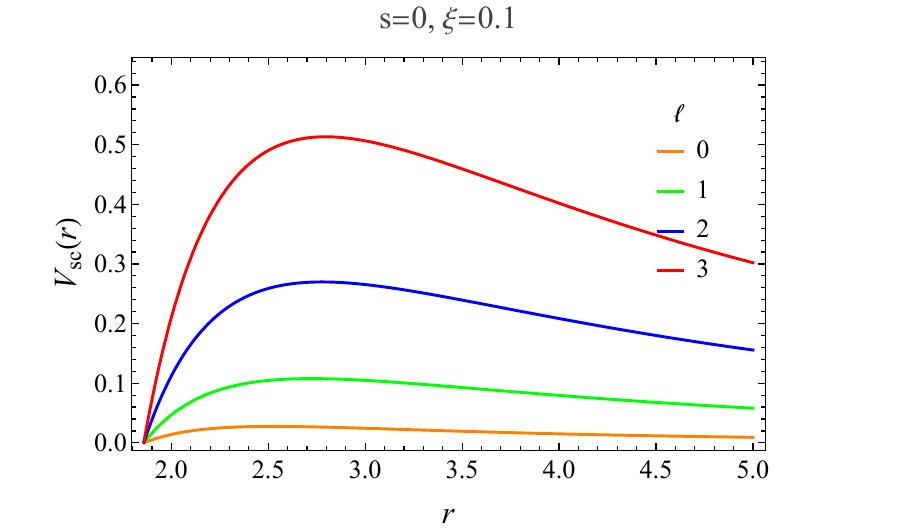}
\caption{\label{Fig:Vsc}Behavior of the scalar effective potential $V_{sc}(r)$ as a function of the radial distance $r$. In the left plot we fix $\ell=0$ and tune $\xi$, while we fix $\xi=0.1$ and tune $\ell$ in the right plot, respectively.}
\end{figure*}

We plot  $V_{sc}(r)$ for different values of the non-minimal coupling parameter $\xi$ and the multipole number $\ell$ in FIG. \ref{Fig:Vsc}. At fixed $\xi=0.1$, the height of the potential barrier increases markedly as $\ell$ increases from $0$ to $3$, similar to the case in Schwarzschild black hole \cite{Kokkotas:1999bd}. This behavior follows from the centrifugal term $f(r)\ell(\ell+1)/r^2$, whose contribution grows with $\ell$. For $\ell=0$, this term vanishes, and the potential is determined by the remaining geometry dependent term. In this case, the curves for different values of $\xi$ lie close to one another: increasing $\xi$ from $0.1$ to $2$ produces only a slight decrease in the barrier height. Thus, within the parameter ranges shown, the barrier height is more sensitive to $\ell$ than to $\xi$. The corresponding effects on the scalar QNM spectrum will be  analyzed in Sec. \ref{sec:QNM}.

\subsection{Electromagnetic perturbation}
We next consider an independent, minimally coupled Maxwell test field. This field is distinct from the non-Abelian YM field supporting the background. Its dynamics are governed by Maxwell’s equations
\begin{eqnarray}\label{Maxwell Eq.}
\nabla_\nu F^{\mu\nu}=\frac{1}{\sqrt{-g}}\partial_\nu(\sqrt{-g}F^{\mu\nu})=0,
\end{eqnarray}
where $F_{\mu\nu}=\partial_\mu A_\nu-\partial_\nu A_\mu$ is the field strength tensor, and $A_\mu$ is the vector potential of the perturbed electromagnetic field. It can be decomposed as \cite{Regge:1957td,Zerilli:1970se}
\begin{eqnarray}
\label{RW decomposition}
A_\mu=e^{-i\omega t}\sum_{\ell ,m}\left[\begin{pmatrix}
0\\
0\\
a^{\ell m}(r) \, \frac{1}{\sin\theta} \, \partial_\varphi Y_{\ell m}\\
-a^{\ell m}(r) \, \sin\theta \, \partial_\theta Y_{\ell m}
\end{pmatrix}
+\begin{pmatrix}
j^{\ell m}(r) \, Y_{\ell m}\\
h^{\ell m}(r) \, Y_{\ell m}\\
k^{\ell m}(r) \, \partial_\theta Y_{\ell m}\\
k^{\ell m}(r) \, \partial_\varphi Y_{\ell m}
\end{pmatrix}\right],
\end{eqnarray}
in which the first column is the axial component with parity $(-1)^{\ell+1}$ while the second term is the polar mode with parity $(-1)^{\ell}$.

Substituting the axial and the polar modes of the electromagnetic perturbations into \eqref{Maxwell Eq.}, we can obtain two decoupled radial equations that can be uniformed into the master equation
\begin{equation}\label{EM eq in r}
f^2(r)\frac{d^2\psi}{dr^2}+f(r)f'(r)\frac{d\psi}{dr}+[\omega^2-V_{EM}(r)]\psi=0,
\end{equation}
with
\begin{eqnarray}
\psi(r)=\left\{ \begin{aligned}
&a^{\ell m}(r) \qquad \qquad \qquad \qquad \text{for axial modes,}\\
\frac{r^2}{\ell(\ell+1)}&\left[i\omega h^{\ell m}(r)+\frac{dj^{\ell m}(r)}{dr}\right] \quad \text{for polar modes.}
\end{aligned} \right.
\end{eqnarray}
Using the tortoise coordinate defined in \eqref{eq:rstar}, the radial equation becomes
\begin{equation}
\label{eq:EM master eq}
\frac{d^2\psi}{dr_*^2}+[\omega^2-V_{EM}(r)]\psi=0,
\end{equation}
and the effective potential is
\begin{equation}\label{eq:Vem}
V_{EM}(r)=\left[1+\frac{r^4}{r^4+2\xi Q^2}(\frac{Q^2}{r^2}-\frac{2M}{r})\right]\frac{\ell(\ell+1)}{r^2},
\end{equation}
where $\ell=1,2,\cdots$ for radiative Maxwell modes.

\begin{figure*}[htbp]
\centering
\includegraphics[height=4cm]{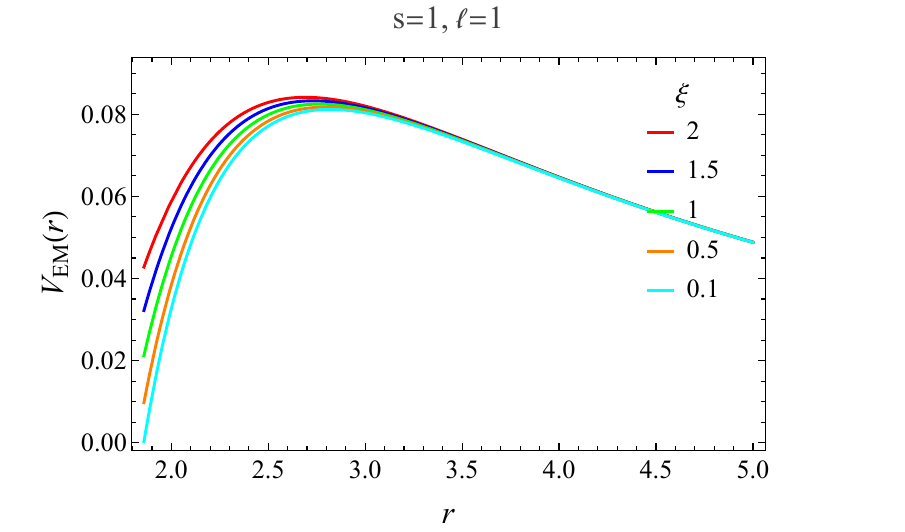}\hspace{0.5cm}
\includegraphics[height=4cm]{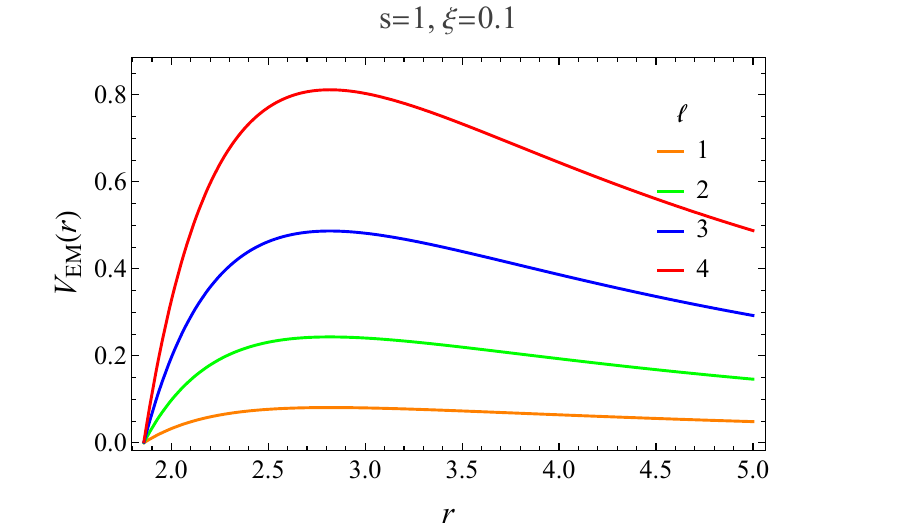}
\caption{\label{fig: V_EM}Behavior of the electromagnetic effective potential $V_{EM}(r)$ as a function of the radial distance $r$. In the left plot We fix $\ell=1$ and tune $\xi$, while we fix $\xi=0.1$ and tune $\ell$ in the  right  plot, respectively.}
\end{figure*}

FIG. \ref{fig: V_EM} shows  $V_{EM}(r)$ for different values of $\xi$ and $\ell$. At fixed $\ell=1$, the five curves lie close to one another, indicating that the potential depends only weakly on $\xi$ within the parameter range considered. Increasing $\xi$ from $0.1$ to $2$ produces only a slight increase in the barrier height. At fixed $\xi=0.1$, the barrier height increases markedly as $\ell$ increases from $1$ to $4$, in qualitative agreement with the behavior observed in the scalar sector.


\subsection{Dirac field perturbation}
We now consider Dirac field perturbations that describe the behaviour of neutral, massless spin-$\frac{1}{2}$ test fermionic fields.
Its equation of motion is
\begin{equation}\label{eq:Dirac eq}
    \Gamma^a [ \partial_a + \frac{1}{4} (\omega_{\mu \nu})_a \Gamma^{\mu \nu} ] \Psi= 0~~~~
 \mathrm{with}~~ \Gamma^{\mu \nu} =\frac{1}{2}[\Gamma^\mu,\Gamma^\nu]~~\mathrm{and}~~
 (\omega_{\mu \nu})_a  =(e_\mu)_b \nabla_a (e_\nu)^b,
\end{equation}
where $\Gamma^a$ are the curved spacetime gamma matrices which are related to their flat spacetime counterparts $\Gamma^{\mu}$ through $
\Gamma^a=(e_\mu)^a\Gamma^{\mu}$, and $(\omega_{\mu\nu})_a$ is the 1-form spin connections. We choose the orthonormal tetrad
$(e^{\mu})_a=\text{Diag}(\sqrt{f},\frac{1}{\sqrt{f}},r,r\sin\theta)$
 whose inverse is denoted by $(e_\mu)^a$ and satisfies $g_{ab}=\eta_{\mu\nu}(e^\mu)_a(e^\nu)_b$ with $\eta_{\mu\nu}=\text{Diag}(-1,1,1,1)$.  Subsequently, the Dirac equation reads
\begin{equation}
\label{eq:DiracExpanded}
    \frac{\Gamma^0}{\sqrt{f}} \frac{\partial\Psi}{\partial t} +
    \Gamma^{1} \sqrt{f} (\frac{\partial}{\partial r} +\frac{1}{r} + \frac{f'}{4f})\Psi+
    \frac{\Gamma^2}{r}(\frac{\partial}{\partial \theta}+\frac{1}{2}\cot\,\theta)\Psi+
    \frac{\Gamma^3}{r \, \sin\theta} \frac{\partial \Psi}{\partial \varphi}=0.
\end{equation}
Following \cite{Cho:2003qe}, we use
\begin{equation}
    \Gamma^0= \begin{pmatrix} -i & 0 \\ 0 & i \end{pmatrix}, \quad
    \Gamma^i= \begin{pmatrix} 0 & -i\sigma^i \\ i\sigma^i & 0 \end{pmatrix}, \quad
    i=1,2,3,
\end{equation}
where $\sigma^i$ denote the standard Pauli matrices.  The Dirac spinor can be decomposed as \cite{Cho:2003qe,Jing:2005dt,Jing:2005pk}
\begin{equation}
    \Psi^{(\pm)}(t,r,\theta,\varphi)= \frac{e^{-i\omega t}}{r f(r)^{\frac{1}{4}}} \begin{pmatrix} iG^{(\pm)}(r) \\ F^{(\pm)}(r) \end{pmatrix} \otimes  \begin{pmatrix} \phi^{(\pm)}_{jm}(\theta,\varphi) \\ \phi^{(\mp)}_{jm}(\theta,\varphi) \end{pmatrix},
\end{equation}
where the spinor spherical harmonics are defined as
\begin{equation}
\begin{split}
        \phi^{(+)}_{jm}=\begin{pmatrix} \sqrt{\frac{j+m}{2j}} Y^{m-1/2}_{l} \\ \sqrt{\frac{j-m}{2j}} Y^{m+1/2}_{l} \end{pmatrix} \qquad \text{for} \quad j=l+\frac{1}{2},\\
    \phi^{(-)}_{jm}=\begin{pmatrix} \sqrt{\frac{j+1-m}{2(j+1)}} Y^{m-1/2}_{l} \\ -\sqrt{\frac{j+1+m}{2(j+1)}} Y^{m+1/2}_{l} \end{pmatrix} \qquad \text{for} \quad j=l-\frac{1}{2}.
    \end{split}
\end{equation}
By substituting the above decomposition into  \eqref{eq:DiracExpanded} and separating the variables, we obtain two decoupled radial master equations
\begin{eqnarray}
    r^2 f \, \partial_{r}(f \, \partial_{r}G^{{(\pm)}}) +(r^2\omega^2-\kappa^2_{\pm}f-\kappa_{\pm}f^{3/2}+\frac{1}{2}\kappa\sqrt{f}f')G^{(\pm)} =0, \\
    r^2 f \, \partial_{r}(f \, \partial_{r}F^{{(\pm)}}) +(r^2\omega^2-\kappa^2_{\pm}f+\kappa_{\pm}f^{3/2}-\frac{1}{2}\kappa\sqrt{f}f')F^{(\pm)} =0,
\end{eqnarray}
where $\kappa$ is related to the total angular momentum by  $\kappa_{\pm}=\mp (j+\frac{1}{2})$ for $j=l \pm 1/2$. In the tortoise coordinate \eqref{eq:rstar},  the above radial equations take the following Schrödinger-like forms
\begin{eqnarray}
\label{eq:Dirac master eq}
    \frac{d^2F^{(\pm)}}{dr^2_*}+[\omega^2-V^{\rom{1}}_{Dirac}]F^{(\pm)}=0, \\
     \frac{d^2G^{(\pm)}}{dr^2_*}+[\omega^2-V^{\rom{2}}_{Dirac}]G^{(\pm)}=0,
\end{eqnarray}
where the Dirac effective potentials are
\begin{eqnarray} \label{eq:Vdirac}
V^{\rom{1}}_{Dirac}=\frac{\sqrt{f}|\kappa_{+}|}{r^2}(|\kappa_{+}|\sqrt{f}+\frac{r f'}{2}-f), \\
    V^{\rom{2}}_{Dirac}=\frac{\sqrt{f}|\kappa_{-}|}{r^2}(|\kappa_{-}|\sqrt{f}-\frac{r f'}{2}+f).
\end{eqnarray}

\begin{figure*}[htbp]
\centering
\includegraphics[height=4cm]{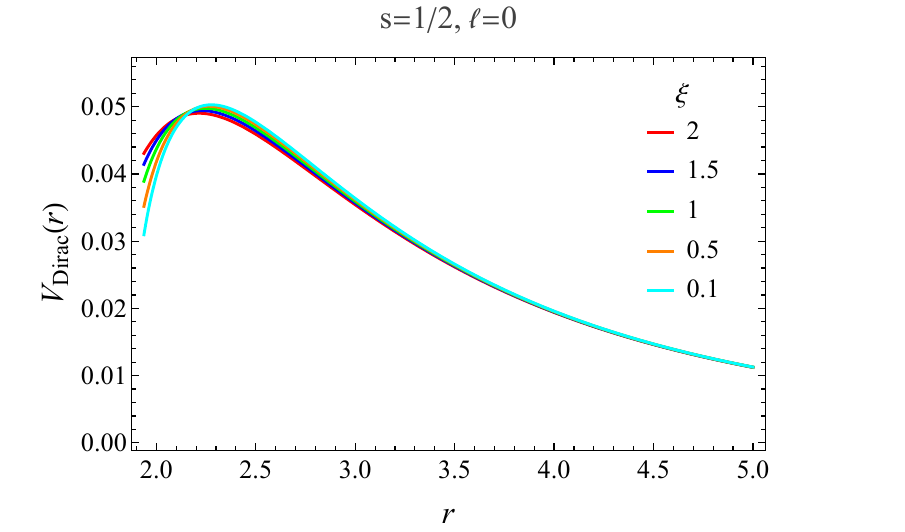}\hspace{0.5cm}
\includegraphics[height=4cm]{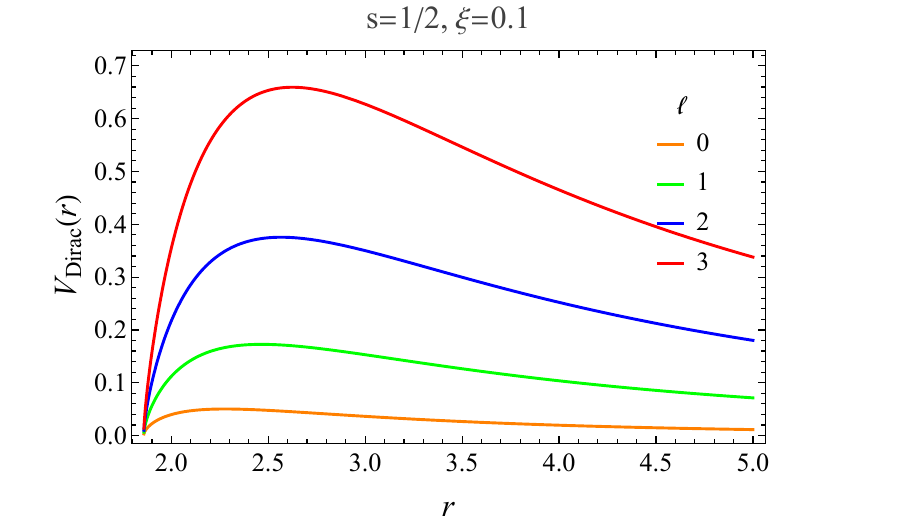}
\caption{\label{fig:V_Dirac}Behavior of the Dirac effective potential $V_{Dirac}^{I}(r)$ as a function of the radial distance $r$. In the left plot, We fix $\ell=0$ and tune $\xi$, while we fix $\xi=0.1$ and tune $\ell$ in the right   plot.}
\end{figure*}

The two  potentials are supersymmetric partners at the same $\kappa$.   With the standard black hole boundary conditions they have the same nonzero QNM spectrum and transmission probabilities \cite{Anderson:1991,Zinhailo:2019rwd}. Therefore, either potential may be adopted in the subsequent analysis without loss of generality. We use  $V_{Dirac}^{I}$ below.
FIG. \ref{fig:V_Dirac} shows  $V_{Dirac}(r)$ for different values of $\xi$ and $\ell$.  At fixed $\ell=0$, increasing $\xi$ from $0.1$ to $2$ slightly lowers the barrier height, as in the scalar case. At fixed $\xi=0.1$, the barrier height increases markedly as $\ell$ increases from $0$ to $3$. Thus, the angular mode dependence is more pronounced than the coupling dependence in the examples shown.

Thus, for the black holes considered here, all three effective potentials exhibit a single barrier outside the event horizon. The differences between the fields enter through the intermediate potential profiles, which can be understood from the analytical formulas of the effective potentials. Unlike the scalar potential, the electromagnetic potential contains no explicit derivative of $f(r)$. At the same $\ell$, the two potentials satisfy $V_{sc}-V_{EM}=f(r)f'(r)/r$, so the scalar potential is higher wherever $f'(r)>0$ outside the horizon. The Dirac potential contains the angular contribution $\kappa^2 f(r)/r^2$ and a derivative term associated with its spinor structure. For the branch used here, it can be written as $V_{Dirac}=W^2+dW/dr_*$, where $W=|\kappa|\sqrt{f(r)}/r$ and $|\kappa|=\ell+1$ in our labeling convention \cite{Cho:2003qe}. The derivative term modifies the barrier profile but does not have a definite sign throughout the exterior region.
For the lowest radiative modes considered here, namely $\ell=0$ for the scalar field, $|\kappa|=1$ for the Dirac field, and $\ell=1$ for the electromagnetic field, the electromagnetic barrier is the highest, followed by the Dirac and scalar barriers over the plotted parameter range. This ordering reflects both the different angular eigenvalues and the field-dependent radial terms. Its implications for the QNM spectra and greybody factors will be analyzed below using the full effective potentials.

\section{Quasi-normal mode frequencies of various perturbations}\label{sec:QNM}

With the effective potentials  established, we now investigate the QNFs of the non-minimal EYM black hole \eqref{eq:metric} under the aforementioned perturbations. The radial solution is purely ingoing ($\sim e^{-i\omega r_*}$) at the event horizon as $r\to r_{+}(r_{*}\to-\infty)$, and purely outgoing ($\sim e^{i\omega r_*}$) at spatial infinity as $r\to \infty(r_*\to +\infty)$ \cite{Konoplya:2011qqh,Berti:2009kk}. In general, the perturbed field equations \eqref{eq:scalar master eq}, \eqref{eq:EM master eq} and \eqref{eq:Dirac master eq} cannot be solved analytically in closed form. In this section, we use the WKB method  \cite{PhysRevD.35.3621,Konoplya:2003ii,Matyjasek:2017psv} and pseudospectral (PS) method~\cite{Zenginoglu:2007jw,PanossoMacedo:2023qzp,Yang:2025hqk} to calculate QNMs frequencies and to cross check the resulting numerical values. To assess the consistency of the two sets of results, we define the relative percentage difference
\begin{eqnarray}
\label{relative error}
\Delta_k= \left| \frac{\text{PS}(\omega_k)-\text{WKB}(\omega_k)}{\text{WKB}(\omega_k)}\right| 100\%, \qquad k=\text{Re}, \text{Im}
\end{eqnarray}
which measures the consistent percentage between the methods.

\subsection{$\xi-$ dependence}
\begin{figure*}[htbp]
\centering
\includegraphics[height=4cm]{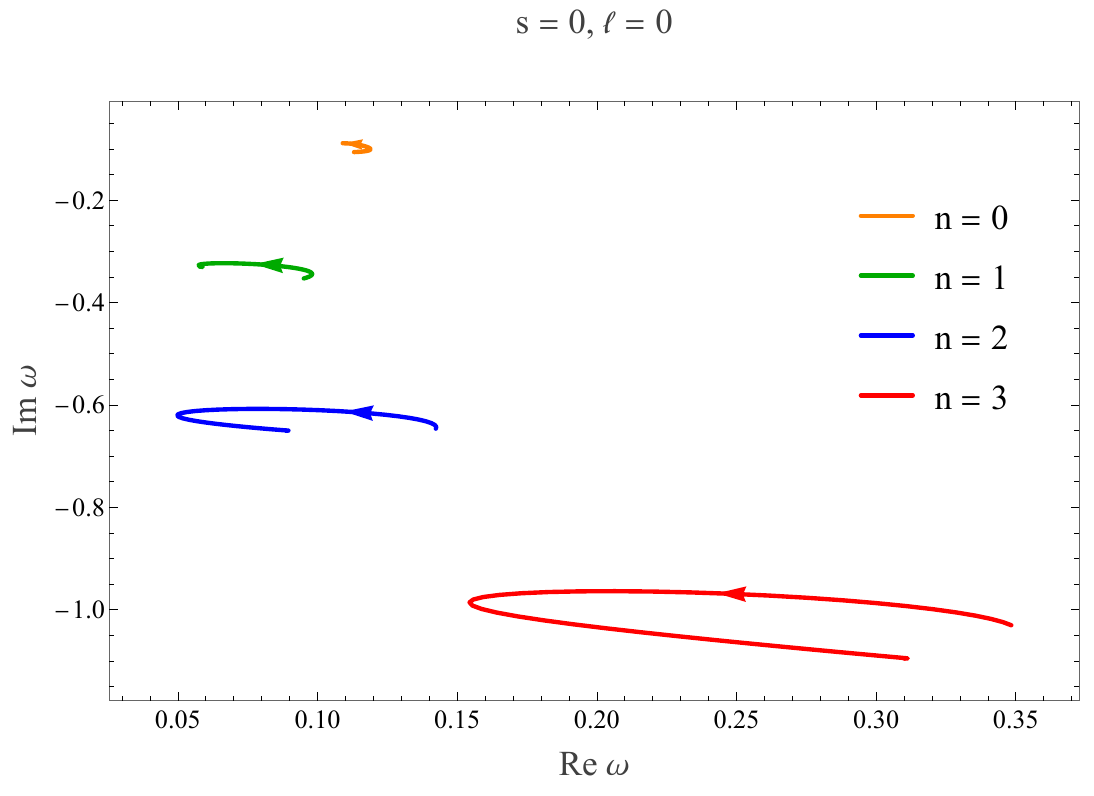}\hspace{0.1cm}
\includegraphics[height=4cm]{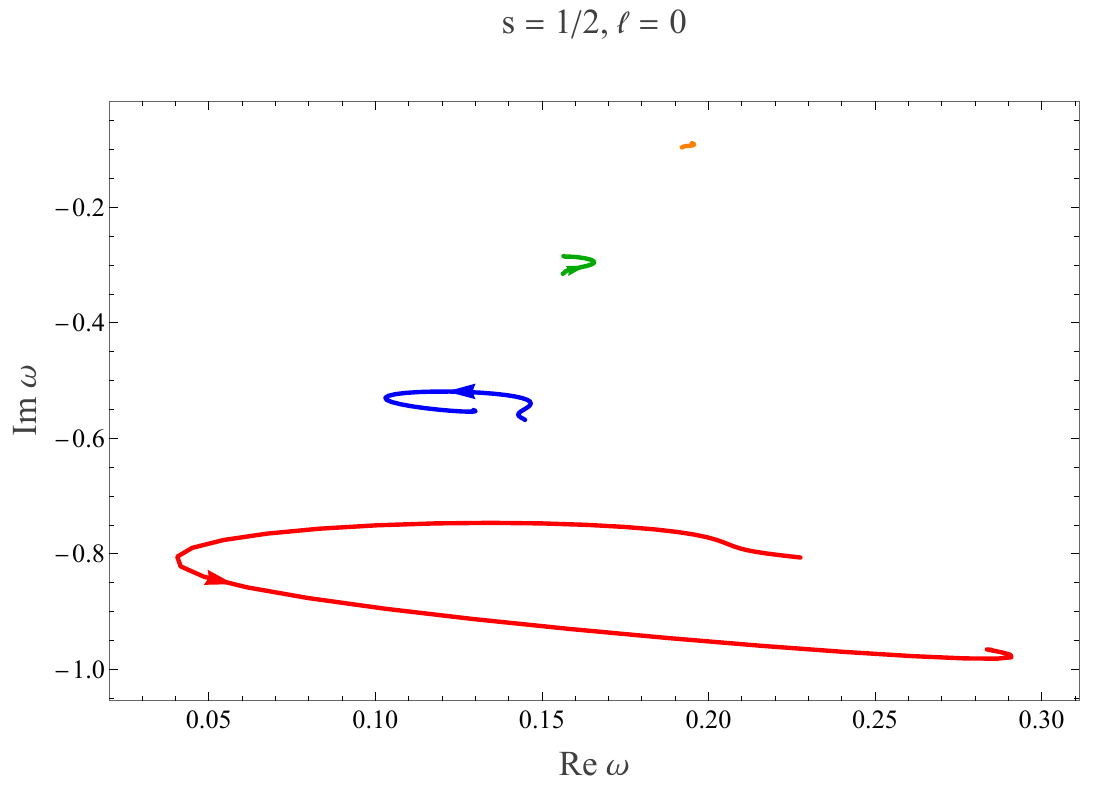}\hspace{0.1cm}
\includegraphics[height=4cm]{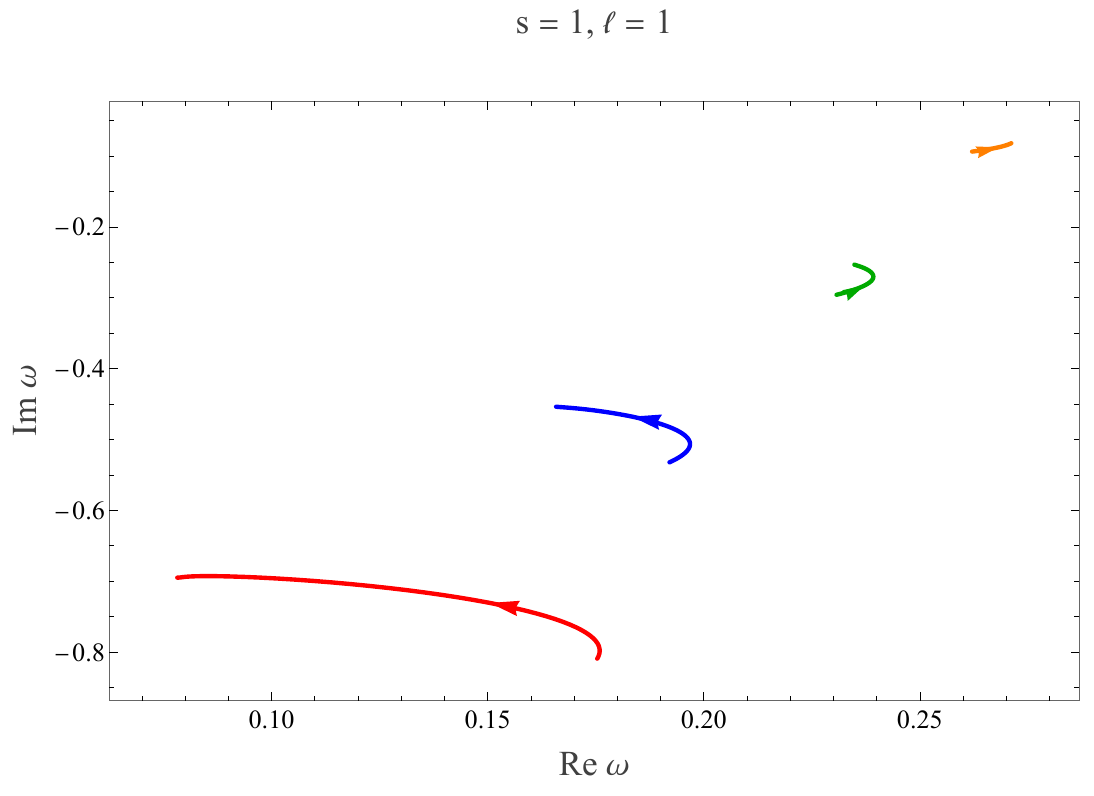}
\\[0.3cm]
\includegraphics[height=4cm]{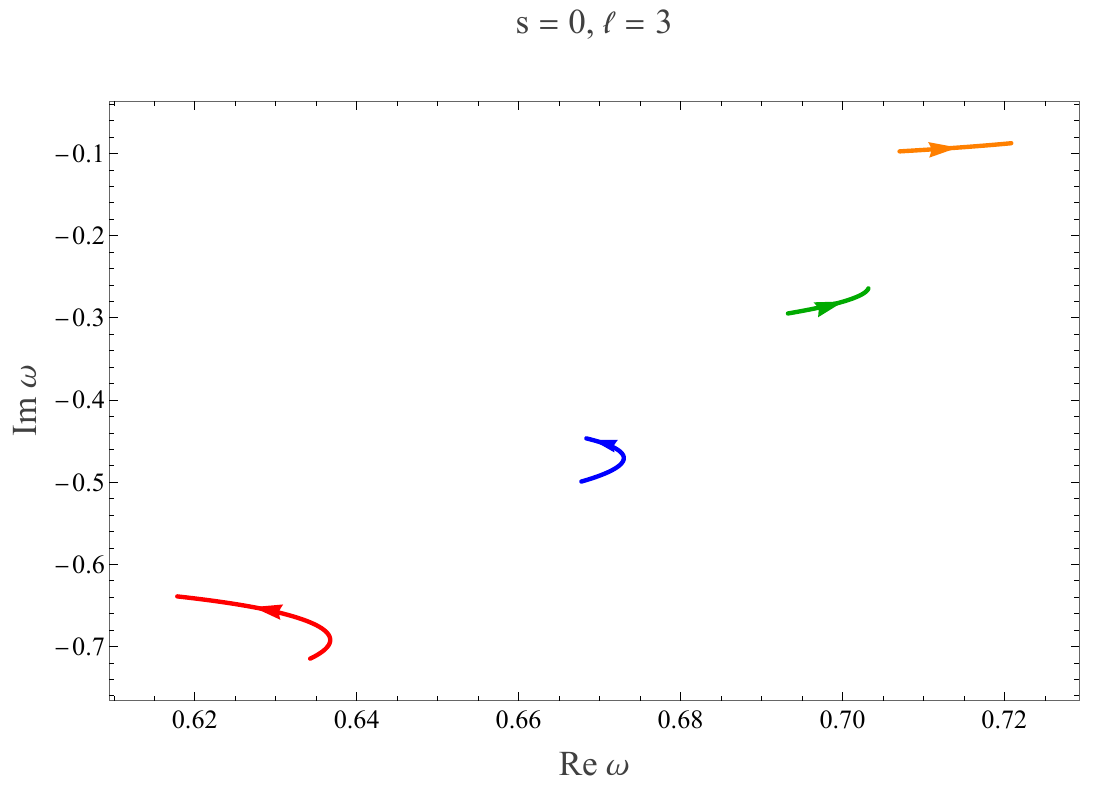}\hspace{0.1cm}
\includegraphics[height=4cm]{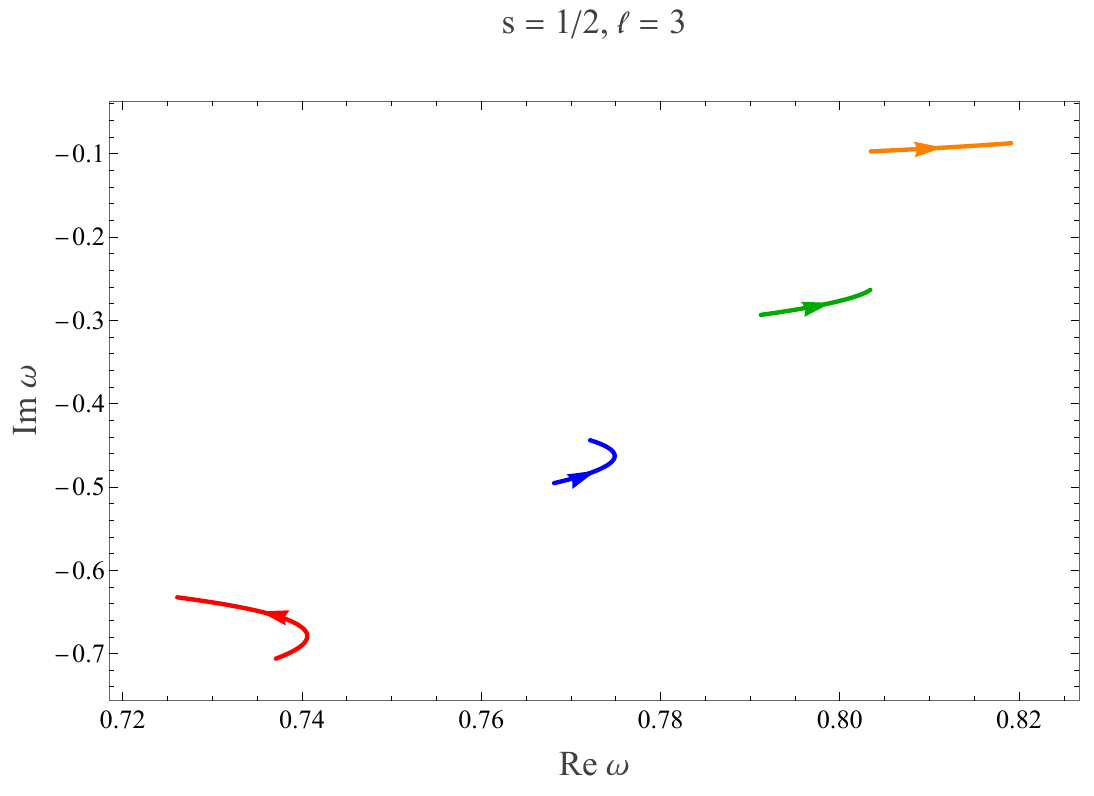}\hspace{0.1cm}
\includegraphics[height=4cm]{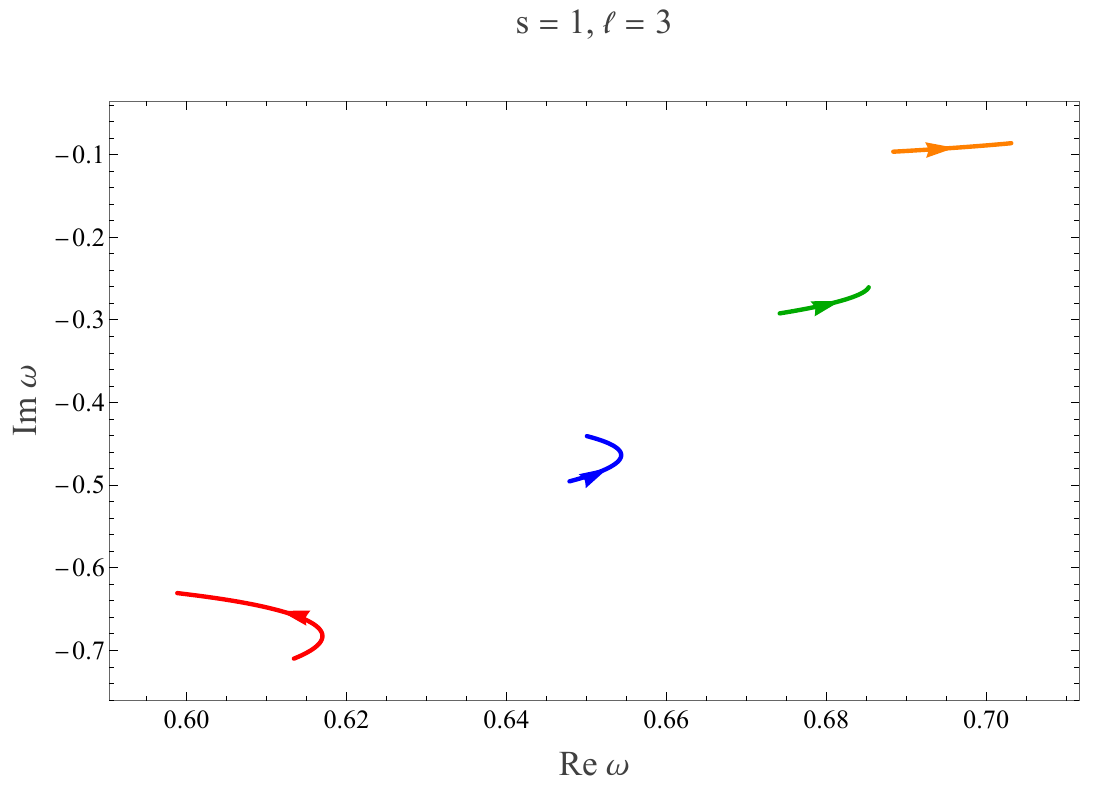}
\caption{\label{matterfieldsQNFl0}Quasi-normal frequencies as a function of $\xi$, which increases from 0 to $\xi_c=2.314$ along the direction of arrows. {\bf Upper panel} is for the lowest multipole number, i.e., $\ell=0$ for scalar $\&$ Dirac fields and $\ell=1$ for EM field. {\bf Bottom panel} is for $\ell=3$. In each plot, the orange, green, blue and red curves are for the overtone $n=0,~1,~2$ and $3$, respectively.}
\end{figure*}

To explore the influence of  $\xi$, we first consider the lowest multipole, i.e.,  $\ell=0$ for scalar and Dirac perturbations and $\ell=1$ for electromagnetic perturbations.
The fundamental frequencies are listed in TABLE \ref{Table:QNM girds varying xi}, while FIG. \ref{matterfieldsQNFl0}  displays the first three overtones for both the lowest multipole and $\ell=3$, respectively. From the table,  we see that the results calculated from WKB and PS method agree well with each other. Moreover, for all cases, the imaginary part of QNM frequencies are always negative. This means that the current black hole is dynamically stable under the test field perturbations.

\begin{table*}[h!]
	\center
	\begin{tabular}{|c|c|c|c|c|} \hline
		\null & \multicolumn{2}{c|}{ scalar field ($s=0$),\quad$\ell=0$} &\multicolumn{2}{c|}{relative error/\%}  \\ \hline
		$\xi$ & Pseudospectral & WKB & Re($\omega$)  & Im($\omega$)  \\ \hline
   2& 0.112777 - 0.087323 i & 0.109916 - 0.088869 i & 2.6029$\%$ & 1.7396$\%$ \\ \hline
   1.8& 0.113914 - 0.091931 i & 0.110863 - 0.089068 i & 2.7520$\%$ & 3.2144$\%$ \\ \hline
   1.4& 0.116445 - 0.095622 i  & 0.114054 - 0.091055 i & 2.0964$\%$ & 5.0156$\%$ \\ \hline
     1& 0.116864 - 0.098903 i & 0.117739 - 0.095956 i & 0.7432$\%$ & 3.0712$\%$\\ \hline
   0.6& 0.116604 - 0.101612 i & 0.118758 - 0.101670 i & 1.8138$\%$ & 0.0570$\%$ \\ \hline
   0.2& 0.116036 - 0.103898 i & 0.114835 - 0.105466 i & 1.0458$\%$ & 1.4867$\%$\\ \hline
   0.1& 0.115868 - 0.104414 i & 0.113061 - 0.105982 i & 2.4827$\%$ & 1.4795$\%$\\ \hline\hline
\null & \multicolumn{2}{c|}{ Dirac field ($s=1/2$),\quad $\ell=0$ } &\multicolumn{2}{c|}{relative error/\%}  \\ \hline
		$\xi$ & Pseudospectral  & WKB & Re($\omega$)  & Im($\omega$)  \\ \hline
  2& 0.191560 - 0.084665 i & 0.195367 - 0.089732 i & 1.9486$\%$ & 5.6468$\%$ \\ \hline
  1.8& 0.194331 - 0.085620 i & 0.195600 - 0.090481 i & 0.6488$\%$ & 5.3724$\%$\\ \hline
  1.4& 0.197444 - 0.089909 i & 0.195357 - 0.092339 i & 1.0683$\%$ & 2.6316$\%$ \\ \hline
  1  & 0.198523 - 0.094314 i & 0.193932 - 0.093584 i & 2.3673$\%$ & 0.7801$\%$\\ \hline
  0.6& 0.198611 - 0.098270 i & 0.192691 - 0.094521 i & 3.0723$\%$ & 3.9663$\%$\\ \hline
  0.2& 0.198170 - 0.101768 i & 0.192173 - 0.095870 i & 3.1206$\%$ & 6.1521$\%$\\ \hline
  0.1& 0.198006 - 0.102579 i & 0.191986 - 0.096171 i & 3.1356$\%$ & 6.6631$\%$\\ \hline\hline
\null & \multicolumn{2}{c|}{ Electromagnetic field ($s=1$),\quad $\ell=1$} &\multicolumn{2}{c|}{relative error/\%}  \\ \hline
		$\xi$ & Pseudospectral  & WKB & Re($\omega$)  & Im($\omega$)  \\ \hline
  2& 0.270393 - 0.083498 i & 0.270391 - 0.083776 i & 0.0007$\%$ & 0.3318$\%$\\ \hline
1.8& 0.268408 - 0.085659 i & 0.269761 - 0.085068 i & 0.5016$\%$ & 0.6947$\%$\\ \hline
1.4& 0.267438 - 0.087534 i & 0.268240 - 0.087510 i & 0.2990$\%$ & 0.0274$\%$\\ \hline
  1& 0.265467 - 0.089569 i & 0.266470 - 0.089690 i & 0.3764$\%$ & 0.1349$\%$ \\ \hline
0.6& 0.263851 - 0.091507 i & 0.264557 - 0.091600 i & 0.2669$\%$ & 0.1015$\%$\\ \hline
0.2& 0.262212 - 0.093183 i & 0.262575 - 0.093274 i & 0.1383$\%$ & 0.0976$\%$\\ \hline
0.1& 0.261804 - 0.093567 i & 0.262075 - 0.093660 i & 0.1034$\%$ & 0.0993$\%$\\ \hline
 \end{tabular}
  \caption{The fundamental ($n=0$) QNM frequencies of lowest $\ell$-mode for various massless field perturbations obtained by WKB method and PS method, and their relative errors.\label{Table:QNM girds varying xi}}
\end{table*}

More interesting properties can be read off from FIG. \ref{matterfieldsQNFl0}, in which we show the QNM frequencies in $(\mathrm{Re}(\omega), \mathrm{Im}(\omega))$ plane. The upper panel is for the modes with lowest multipole number for each test field.  For the scalar perturbations (upper-left plot), the variation of the QNM frequencies with  $\xi$ becomes more pronounced at higher overtone numbers. For the fundamental mode ($n=0$), $\mathrm{Re}(\omega)$ initially increases slightly and then decreases as $\xi$ grows. The first overtone ($n=1$) exhibits similar behavior. For $n=2,3$, however, $\mathrm{Re}(\omega)$ first decreases, reaches a minimum at an intermediate value of $\xi$, and then increases, with a larger variation for $n=3$. The imaginary parts also distinguish the lower modes from the higher overtones. For $n=0,1$, $\mathrm{Im}(\omega)$ generally becomes less negative as $\xi$ increases, indicating weaker damping. For $n=2,3$, it first increases and then decreases, so the damping rate initially decreases before increasing again at larger $\xi$.
For  Dirac perturbations (upper-middle plot), the real parts of the fundamental mode and the first overtone vary smoothly with $\xi$, while the higher overtones exhibit more pronounced non-monotonic behavior. In particular, $\mathrm{Re}(\omega)$ for $n=2,3$ first decreases and then increases, and this behavior repeats again, indicating with multi turning points. The imaginary parts of the lower modes become less negative as $\xi$ increases. For the higher overtones, they instead reach a maximum and then decrease, and a further increasing behavior is obvious for $n=3$. This corresponds to transitions between weaker to stronger damping. While for electromagnetic perturbations (upper-right plot), $\mathrm{Re}(\omega)$ varies smoothly with $\xi$ for $n=0$, whereas non-monotonic behavior appears for $n=1,2,3$. The imaginary parts show a more uniform trend across the displayed modes, generally becoming less negative as $\xi$ increases. The electromagnetic modes therefore become longer lived over the parameter range considered, without the reversal in the damping trend observed for the higher scalar and Dirac overtones.

The different responses of these fields reflect the distinct angular and radial terms in their effective potentials. For the scalar mode with $\ell=0$, the centrifugal term vanishes, leaving $V_{sc}=ff'/r$, so both the metric function and its radial derivative enter the potential directly. Nevertheless, the changes in the barrier height alone do not determine the response of the full QNM spectrum. The stronger variation of higher overtones is qualitatively consistent with the phenomenon known as overtone outburst \cite{Konoplya:2022hll}, in which the small near-horizon deformations produce much larger relative frequency shifts for the overtones than for the fundamental mode. The behavior found here suggests a similar sensitivity to the coupling-induced deformation, while the distinct damping trends show that its spectral consequences depend on the field and overtone considered. Higher overtones can therefore provide information about the background geometry that is less apparent in the fundamental mode. However, these differences in various perturbations become less apparent at $\ell=3$ as shown in the bottom panel of FIG. \ref{matterfieldsQNFl0}, from which we see that the QNM frequencies
with the same node for all case behave similarly. This is reasonable because  the effective potentials for those perturbations coincide for sufficicently large enough $\ell$. Next, we shall study the effect of $\ell$ on the QNM frequencies more carefully.

\subsection{$\ell-$ dependence}
\begin{figure*}[htbp]
 \centering
\includegraphics[height=4cm]{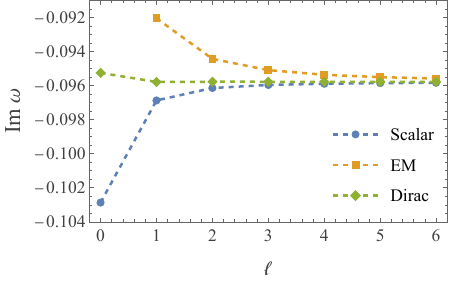}\hspace{0.5cm}
\includegraphics[height=4cm]{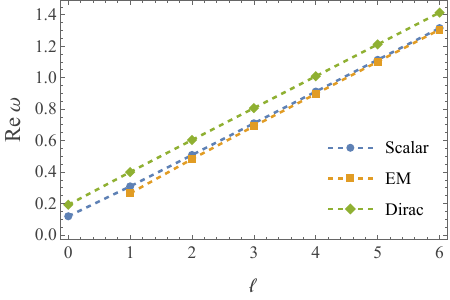}
\caption{\label{matterfieldsQNFl1} Fundamental QNM frequencies as a function of the multipole number for various test fields.}
\end{figure*}

In this subsection, we fix $\xi=1/2$ to examine the fundamental mode as $\ell$ increases.  The results are given in TABLE \ref{Table:QNM grids varying l} and FIG. \ref{matterfieldsQNFl1}.

\begin{table*}[h!]
	\center
	\begin{tabular}{|c|c|c|c|c|} \hline
\null & \multicolumn{2}{c|}{ scalar field ($s=0$)} &\multicolumn{2}{c|}{relative error/\%}  \\ \hline
		$\ell$ & PS  & WKB & Re($\omega$)  & Im($\omega$)  \\ \hline
  0& 0.116482 - 0.102219 i & 0.118276 - 0.102868 i & 1.5168$\%$ & 0.6309$\%$\\ \hline
  1& 0.308207 - 0.096925 i & 0.308241 - 0.096865 i & 0.0110$\%$ & 0.0619$\%$\\ \hline
  2& 0.507968 - 0.096268 i & 0.508094 - 0.096136 i & 0.0248$\%$ & 0.1373$\%$\\ \hline
  3& 0.709380 - 0.095882 i & 0.709418 - 0.095956 i & 0.0054$\%$ & 0.0771$\%$\\ \hline
  4& 0.911171 - 0.095907 i & 0.911131 - 0.095885 i & 0.0044$\%$ & 0.0229$\%$\\ \hline
  5& 1.112976 - 0.095846 i & 1.113007 - 0.095849 i & 0.0028$\%$ & 0.0031$\%$\\ \hline\hline
\null & \multicolumn{2}{c|}{ Dirac field ($s=1/2$)} &\multicolumn{2}{c|}{relative error/\%}  \\ \hline
		$\ell$ & PS  & WKB & Re($\omega$)  & Im($\omega$)  \\ \hline
  0& 0.198540 - 0.099185 i & 0.191192 - 0.095246 i & 3.8433$\%$ & 4.1356$\%$\\ \hline
  1& 0.399388 - 0.094462 i & 0.399513 - 0.095779 i & 0.03129$\%$ & 1.3750$\%$\\ \hline
  2& 0.602846 - 0.095881 i & 0.603163 - 0.095767 i & 0.05256$\%$ & 0.1190$\%$\\ \hline
  3& 0.806151 - 0.095845 i & 0.806107 - 0.095768 i & 0.0055$\%$ & 0.0804$\%$\\ \hline
  4& 1.008772 - 0.095752 i & 1.008744 - 0.095771 i & 0.0028$\%$ & 0.0198$\%$\\ \hline
  5& 1.211216 - 0.095758 i & 1.211222 - 0.095773 i & 0.0005$\%$ & 0.0157$\%$\\ \hline\hline
\null & \multicolumn{2}{c|}{  Electromagnetic field ($s=1$)} &\multicolumn{2}{c|}{relative error/\%}  \\ \hline
		$\ell$ & PS  & WKB & Re($\omega$)  & Im($\omega$)  \\ \hline
  1& 0.263441 - 0.091949 i & 0.264066 - 0.092039 i & 0.2367$\%$ & 0.09779$\%$\\ \hline
  2& 0.481984 - 0.094423 i & 0.482025 - 0.094421 i & 0.0085$\%$ & 0.0021$\%$\\ \hline
  3& 0.690888 - 0.095086 i & 0.690895 - 0.095085 i & 0.0010$\%$ & 0.0011$\%$\\ \hline
  4& 0.896753 - 0.095359 i & 0.896754 - 0.095359 i & 0.0001$\%$ & $\mathcal{O}^{-6}$\\ \hline
  5& 1.101256 - 0.095497 i & 1.101257 - 0.095497 i & 0.0001$\%$ & $\mathcal{O}^{-6}$\\ \hline
  6& 1.305032 - 0.095577 i & 1.305032 - 0.095577 i & $\mathcal{O}^{-6}$ & $\mathcal{O}^{-6}$\\ \hline
	\end{tabular}
\caption{The fundamental ($n=0$) QNMs of various test fields with different angular momentums obtained by WKB method and PS method and their relative errors. Here we fix $\xi=1/2$.
\label{Table:QNM grids varying l}}
 \end{table*}

From the right panel of FIG. \ref{matterfieldsQNFl1}, it can be observed that $\mathrm{Re}(\omega)$ increases monotonically with $\ell$ for all three perturbation fields. Although the oscillation frequencies differ at low multipoles, their real parts exhibit a common leading linear growth at large $\ell$. The damping rates, however, approach the large multipole regime in different ways. While in the left panel, as $\ell$ increases, $|\mathrm{Im}(\omega)|$ decreases for scalar perturbations and increases for electromagnetic perturbations, whereas the Dirac damping rate remains nearly unchanged. 
Similar phenomena also occur in Schwarzschild black hole \cite{Zhang:2006hh} and Horndeski hairy black hole studied found in \cite{Yang:2023lcm,Yang:2024cjf}. Thus, the angular dependence of the oscillation frequencies is similar across the three sectors, while the finite-$\ell$ corrections to their damping rates remain field dependent.
These trends again can be understood from the structure of the effective potentials. At large $\ell$, the dominant angular terms share the same radial dependence, $f(r)/r^2$, while the remaining field dependent contributions become subleading. Consequently, for fixed overtone number, the three sectors share a common leading eikonal behavior governed by the unstable circular null orbit \cite{Cardoso:2008bp}. Their oscillation frequencies grow with the appropriate angular eigenvalues, and their damping rates approach the same limiting value. The relative differences between the oscillation frequencies therefore diminish at large $\ell$, although angular offsets between the bosonic and Dirac sectors must be retained when comparing the frequencies themselves. This correspondence will be examined in the following subsection.

\subsection{QNM frequencies in the eikonal limit}
In the eikonal limit ($\ell\gg1$), the photon sphere branch of the test field QNMs is related to unstable circular null geodesics \cite{Cardoso:2008bp}. We denote the photon sphere radius by $r_{ph}$, which for the current  black hole is determined by
\begin{eqnarray}
    2f(r_{ph})-r_{ph}f'(r_{ph})=0.
\end{eqnarray}
The orbital angular velocity $\Omega_{ph}$ and Lyapunov exponent $\lambda_{ph}$ of the unstable circular null geodesic are given by
\begin{eqnarray}\label{eq:OmegaLambda}
  \Omega_{ph}=\sqrt{\frac{f(r_{_{ph}})}{r^2_{_{ph}}}},~~~\text{and}~~~\lambda_{ph}&=&\sqrt{-\frac{r^2_{ph}}{2f(r_{ph})}\left(\frac{d^2}{dr^2_*}\frac{f(r)}{r^2}\right)_{r=r_{ph}}}.
\end{eqnarray}

\begin{table*}[h!]
	\center
	\begin{tabular}{|c|c|c|c|c|} \hline
\multicolumn{2}{|c|}{\null} & PS method  &  WKB method & Eq.\eqref{eq:eikonalQNM}
\\\hline
\multirow{2}{*}{scalar}  & $\xi=0.5$ & 20.315882 - 0.095778 i & 20.315882 - 0.095778 i & 20.315817 - 0.095778 i  \\ \cline{2-5}
     ~& $\xi=1$ & 20.407358 - 0.093898 i & 20.407358 - 0.093898 i & 20.407298 - 0.093898 i  \\\hline
\multirow{2}*{EM} & $\xi=0.5$ & 20.315240 - 0.095777 i & 20.315240 - 0.095777 i & 20.315817 - 0.095778 i  \\\cline{2-5}
     ~ & $\xi=1$    & 20.406722 - 0.093897 i & 20.406722 - 0.093897 i & 20.407298 - 0.093898 i  \\\hline
\multirow{2}*{Dirac} & $\xi=0.5$ & 20.416794 - 0.095780 i & 20.416792 - 0.095778 i   & 20.416891 - 0.095778 i  \\\cline{2-5}
      ~ & $\xi=1$    & 20.508720 - 0.093898 i & 20.508720 - 0.093898 i   & 20.508827 - 0.093898 i  \\\hline
	\end{tabular}
 \caption{QNM frequencies with  $\ell=100$ for various perturbing fields obtained via different methods.
 \label{Table:eikonalQNMtable}}
\end{table*}

\begin{figure*}[htbp]
\centering
\includegraphics[height=4cm]{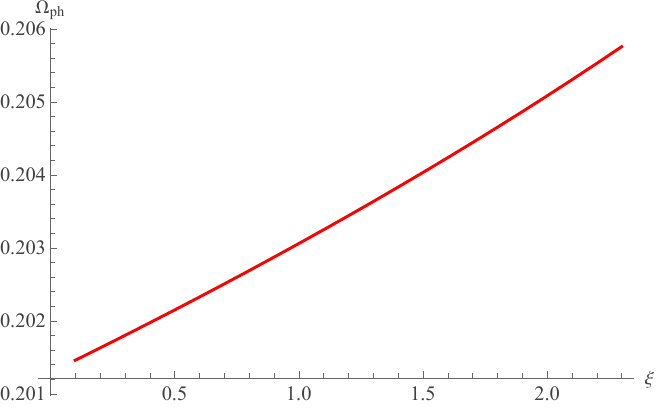}\hspace{1cm}
\includegraphics[height=4cm]{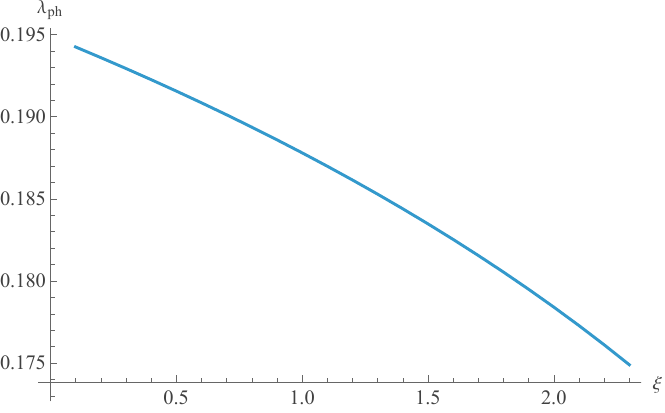}
\caption{\label{eikonalQNMfig} The angular velocity and the Lyapunov exponent with the formulas \eqref{eq:OmegaLambda} as functions.}
\end{figure*}

For a large but finite $\ell$, we define 
\begin{equation}
L_s=
\begin{cases}
\ell+\dfrac{1}{2},~~~ s=0,1,\\
|\kappa|=\ell+1, ~~~ s=\dfrac{1}{2}.
\end{cases}
\end{equation}
The bosonic angular offset follows from $\ell(\ell+1)=(\ell+1/2)^2-1/4$, while the Dirac angular dependence is determined by $|\kappa|$ \cite{Dolan:2009nk,Yang:2023lcm}.
Then, the photon sphere branch of the massless test field QNMs for fixed overtone number is related to the unstable circular null geodesic as 
\begin{equation}\label{eq:eikonalQNM}
\omega_{sn}
=L_s\Omega_{\rm ph}
-i\left(n+\frac{1}{2}\right)\lambda_{\rm ph}.
\qquad L_s\gg1,
\end{equation}
This correspondence follows from the  leading structure of the effective potentials, $V_{\rm eff}(r)\sim L_s^2f(r)/r^2$. The field-dependent terms contribute at subleading orders, so all three sectors share the limits {$\lim\limits_{\ell \to \infty} \dfrac{\mathrm{Re}(\omega_{s n})}{L_s} = \Omega_{\mathrm{ph}}$ and $\lim\limits_{\ell \to \infty} \mathrm{Im}(\omega_{sn}) = -\left(n + \dfrac{1}{2}\right)\lambda_{\mathrm{ph}}$}. 
To assess the accuracy of this approximation, we compare it with the numerical results for the fundamental modes at $\ell=100$ in Table \ref{Table:eikonalQNMtable}. The small residual differences are consistent with subleading corrections omitted from the geometric-optics approximation. These results support the photon sphere description of the large $\ell$ test field modes in the current black hole.

In order to see the effect of $\xi$, we plot $\Omega_{\rm ph}$ and  $\lambda_{\rm ph}$ as functions of $\xi$ in FIG. \ref{eikonalQNMfig}, which shows how these two geometric quantities depend on $\xi$. Within the range considered, $\Omega_{\rm ph}$ increases monotonically with $\xi$, whereas $\lambda_{\rm ph}$ decreases. At fixed $L_s$ and $n$, the former leads to a higher oscillation frequency through $\mathrm{Re}(\omega)\simeq L_s\Omega_{\rm ph}$. The latter indicates a longer instability timescale for the null orbit and a smaller QNM damping rate, $|\mathrm{Im}(\omega)|\simeq(n+1/2)\lambda_{\rm ph}$. Increasing $\xi$ therefore produces faster oscillations and slower decay for the eikonal modes. The different test fields share this leading response because it is determined by the same background geometry.

\section{Greybody factors and Hawking radiation}\label{sec:Hawking Radiation}

Hawking showed that black holes emit radiation with a thermal occupation factor determined by their Hawking temperature \cite{Hawking:1975vcx}. The spectrum measured at infinity is further modified by scattering through the effective potential outside the event horizon. The corresponding transmission probability, known as the greybody factor, depends on the frequency, angular mode and type of the emitted field \cite{Page1976}. In this section, we first calculate the greybody factors for the three massless test fields and then examine how these transmission probabilities, together with the Hawking temperature, determine their emission spectra.

\subsection{Greybody factor}

We utilize the sixth-order WKB approximation  to calculate greybody factors for the scalar, electromagnetic and Dirac fields perturbations.
For a real frequency $\omega>0$ and the time dependence $e^{-i\omega t}$, we solve the master equations \eqref{eq:scalar master eq},  \eqref{eq:EM master eq} and \eqref{eq:Dirac master eq} subject to scattering boundary conditions. These describe a wave incident from infinity, with a reflected component at infinity and a purely ingoing component at the event horizon,
\begin{equation}
\begin{split}
    \Psi_{\ell}&=T_{\ell}\, e^{-i\omega r_*},\qquad \quad r_*\rightarrow -\infty,\\
     \Psi_{\ell}&=e^{-i\omega r_*}+R_{\ell}\, e^{i\omega r_*}, \, \qquad \quad r_*\rightarrow +\infty.\\
\end{split}
\end{equation}
Here, $T_{\ell}$ and $R_{\ell}$ are the transmission and reflection amplitudes, respectively, with the incident amplitude normalized to unity. For the test fields considered here, flux conservation gives $|T_{\ell}|^2+|R_{\ell}|^2=1$.
The greybody factor is therefore \cite{Schutz:1985km,PhysRevD.35.3621}
\begin{equation}
|A_{\ell}|^2=|T_{\ell}|^2=1-|R_{\ell}|^2.
\end{equation}
It represents the absorption probability for a wave incident from infinity. By reciprocity, the same transmission probability enters the outgoing Hawking flux.

Within the WKB approximation, the reflection and transmission probabilities take the form
\begin{equation}
|R_{\ell}|^2\simeq
\frac{1}{1+e^{-2i\pi\mathcal{K}}},
\qquad
|A_{\ell}|^2\simeq
\frac{1}{1+e^{2i\pi\mathcal{K}}},
\end{equation}
where $\mathcal{K}$ is obtained from
\begin{equation}
\mathcal{K}
-i\frac{\omega^2-V_0}{\sqrt{-2V_0^{(2)}}}
+\sum_{j=2}^{6}\Lambda_j(\mathcal{K})=0.
\end{equation}
Here, $V_0=V(r_0)$ denotes the maximum of the effective potential for the field under consideration, and
$V_0^{(p)}\equiv\left.\frac{d^pV}{dr_*^p}\right|_{r=r_0}$.
The correction terms $\Lambda_j$ involve derivatives of the potential up to order $2j$. The relevant solution for $\mathcal{K}$ is the branch connected to the leading-order expression
$\mathcal{K}=i(\omega^2-V_0)/\sqrt{-2V_0^{(2)}}$.
Since this WKB expansion is constructed around the potential maximum, its accuracy can deteriorate at sufficiently low frequencies, particularly for the scalar $\ell=0$ mode \cite{Konoplya:2019hlu}. Thus, we introduce a lower frequency cutoff $\omega_{\min}$ in the numerical evaluation. The cutoff avoids extrapolating the approximation to $\omega\rightarrow0$, but its influence must also be considered when evaluating frequency integrated emission rates.

\begin{figure*}[htbp]
\centering
\includegraphics[height=4cm]{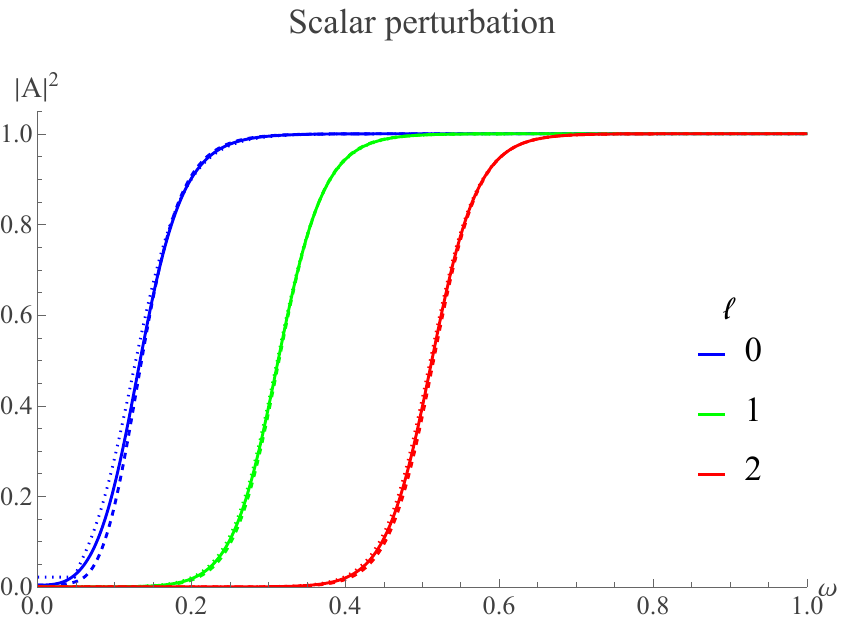}\hspace{0.1cm}
\includegraphics[height=4cm]{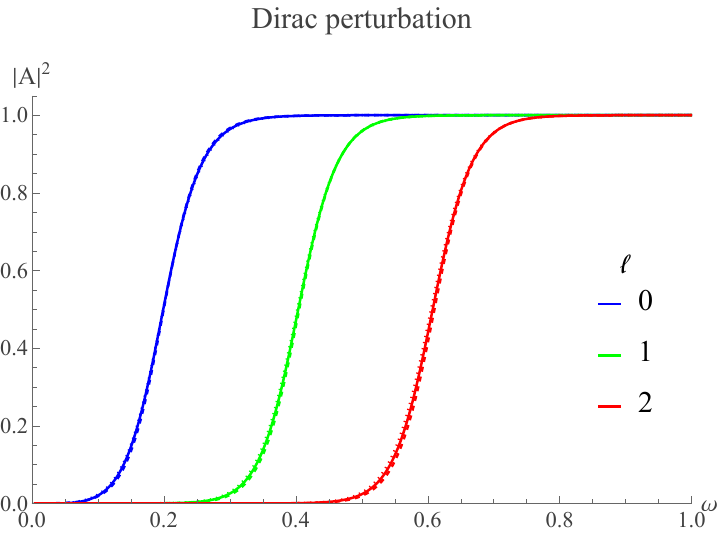}\hspace{0.1cm}
\includegraphics[height=4cm]{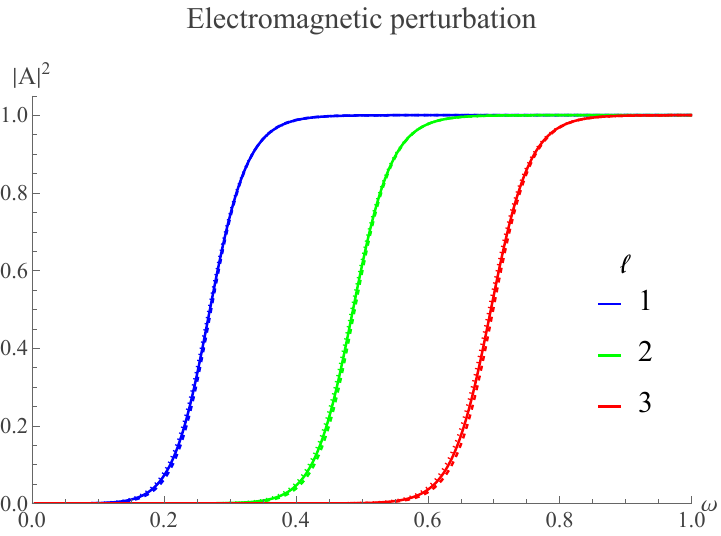}
\caption{\label{graybodyA}Greybody factors for  {emitted} particles with spin $0$ (left), $1/2$ (middle) and $1$ (right). In each plot, we distinguish the curves with different $\ell$ by colors, while curves with same color indicate the graybody factor with $\xi=1/2 $ (dotted), $\xi=1$ (solid), $\xi=3/2$ (dashed).}
\end{figure*}

FIG.~\ref{graybodyA} shows the greybody factors as functions of $\omega$ for different $\ell$ and  $\xi$. The three perturbation sectors exhibit the same general transition from strong reflection at low frequencies to nearly complete transmission at high frequencies. As $\omega$ increases, the transmission probability rises rapidly and approaches unity for each fixed partial wave.
Within each sector, increasing $\ell$ shifts this transition towards higher frequencies, consistent with the increasing angular contribution to the effective potential. Among the lowest radiative partial waves, the transition occurs at progressively higher frequencies for the scalar field with $\ell=0$, the Dirac field with $\ell=0$ and $|\kappa|=1$, and the electromagnetic field with $\ell=1$. This ordering depends on both the field equations and the angular modes chosen for comparison.

Compared with the pronounced dependence on $\ell$, the effect of $\xi$ is relatively weak. The curves separate mainly in the frequency interval where the transmission probability rises rapidly, while their differences diminish as the greybody factors approach unity.
For the scalar field, the response to $\xi$ depends on the multipole number. In the lowest partial wave, $\ell=0$, the centrifugal term vanishes and the effective potential reduces to $V_{sc}=f(r)f'(r)/r$. The corresponding greybody factor curves do not maintain a uniform ordering throughout the frequency range: increasing $\xi$ can slightly enhance transmission in some intervals and suppress it in others. For the higher scalar multipoles, larger $\xi$ generally reduces the transmission probability within the main transition region. The electromagnetic and Dirac curves also generally indicate a small suppression of transmission with increasing $\xi$ in this region. These changes should be interpreted through the full effective potential as a function of $r_*$. Transmission depends on the width and shape of the barrier as well as its height. In particular, the lowest Dirac potential has a slightly lower maximum at larger $\xi$, as shown in FIG.~\ref{fig:V_Dirac}.

\subsection{Energy emission rate}
Once the greybody factors are obtained, the Hawking emission spectrum can be calculated. We adopt a quasistationary approximation in which the background parameters and the Hawking temperature are held fixed when evaluating each instantaneous spectrum. This approximation describes an isolated, slowly evolving black hole.
For a field of spin $s$, the energy spectrum measured at infinity is \cite{Hawking:1975vcx,Page1976}.
\begin{equation}
\label{eq}
\frac{\partial^2E}{\partial t \partial\omega}
=\sum_{\ell}\mathcal{P}_{s\ell}(\omega), ~~~\text{with}~~~\mathcal{P}_{s\ell}(\omega)=\frac{N_{s\ell}}{2\pi}
\frac{\omega |A_{s\ell}(\omega)|^2}
{e^{\omega/T_H}\pm1}.
\end{equation}
The minus sign applies to bosons and the plus sign to fermions.  For the non-minimal EYM black hole \eqref{eq:metric}, the Hawking temperature is
\begin{equation}
\label{Eq}
T_H=\frac{f'(r_h)}{4\pi}
=\frac{r_h^4-Q^2r_h^2-6\xi Q^2}
{4\pi r_h\left(r_h^4+2\xi Q^2\right)},
\end{equation}
where $r_h$ is the event horizon determined by $f(r_h)=0$.  And  the multiplicity factors are given by
\begin{equation}
N_{s\ell}=
\begin{cases}
2\ell+1,~~\text{scalar field},\ \\
8(\ell+1),~~ \text{Dirac field},\ \\
2(2\ell+1),~~\text{electromagnetic field}.
\end{cases}
\end{equation}

\begin{figure*}[htbp]
\centering
\includegraphics[height=4cm]{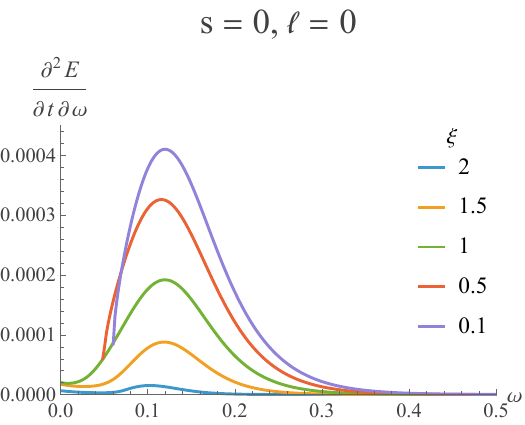}\hspace{0.1cm}
\includegraphics[height=4cm]{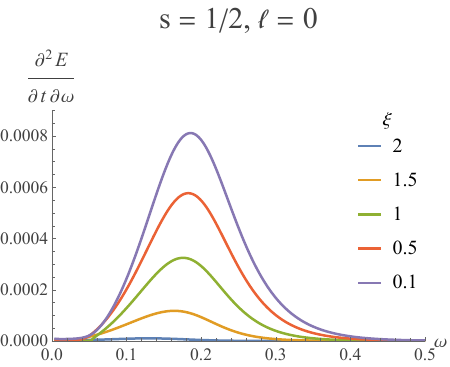}\hspace{0.1cm}
\includegraphics[height=4cm]{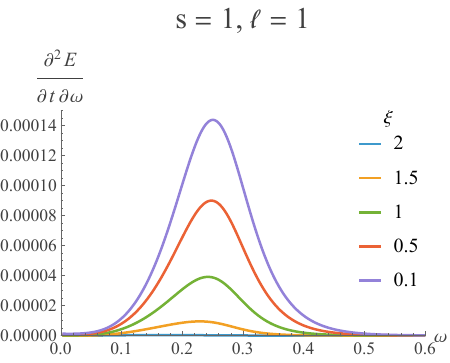}\\
\includegraphics[height=4cm]{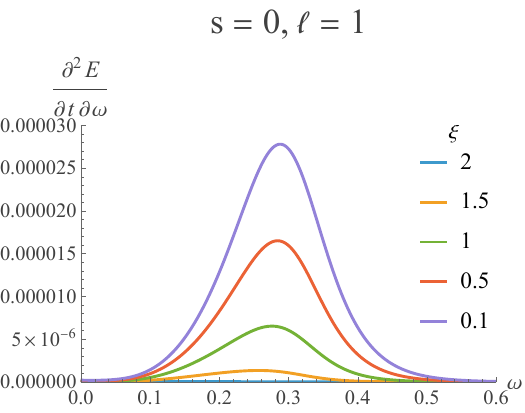}\hspace{0.1cm}
\includegraphics[height=4cm]{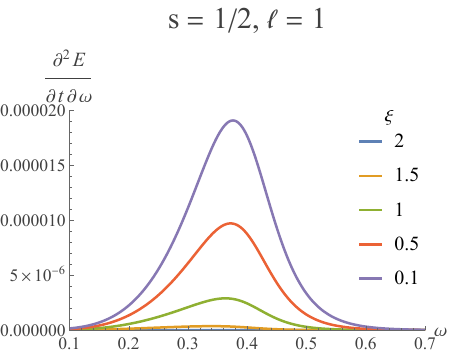}\hspace{0.1cm}
\includegraphics[height=4cm]{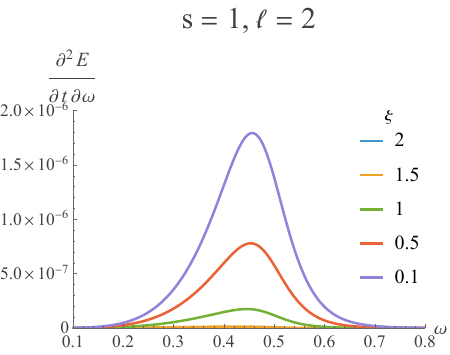}
\caption{\label{EERperw total}Energy emission rate as a function of frequency for various fields with different spins around the non-minimal EYM black hole. The upper panels show the fields with the corresponding lowest $\ell$ while the bottom panel  is for their second lowest $\ell$.}
\end{figure*}

FIG. \ref{EERperw total} displays the energy emission rate for different values of $\xi$. Each spectrum has a pronounced maximum, reflecting the frequency dependence of both the greybody factor and the thermal occupation factor. At low frequencies, the small transmission probability suppresses the outgoing flux. As the frequency increases, transmission becomes more efficient and the spectrum rises. At sufficiently high frequencies, the thermal exponential dominates and causes the spectrum to decay.
Increasing $\xi$ substantially lowers the spectral peaks in all the displayed channels. This change is more pronounced than the corresponding variation of the greybody factors. For the parameter range considered here, the Hawking temperature decreases with increasing $\xi$, which then reduces the thermal occupation factor at every fixed positive frequency, providing a common source of suppression for all three fields.

For the electromagnetic and Dirac fields, as well as the higher scalar multipoles, this thermal suppression is accompanied by a generally reduced transmission probability in the transition region. The lowest scalar partial wave requires a more careful interpretation because its greybody factor can increase slightly over some frequency intervals. Such a local enhancement does not necessarily increase the emission spectrum or its frequency integral, since it can be outweighed by the reduction in the thermal occupation factor. The greybody factors and the Hawking temperature must therefore be considered together when interpreting the dependence of the radiation on $\xi$ \cite{KonoplyaZinhailo2019}.

By comparing the plots in each column in FIG. \ref{EERperw total}, we see that $\ell$ also affects both the height and the position of the spectral peak. The higher mode has a smaller peak at a larger frequency. Increasing $\ell$ shifts appreciable transmission towards higher frequencies, where the thermal occupation factor is smaller. The multiplicity $N_{s\ell}$ increases with $\ell$, but this increase is still too low to compensate for the suppression of transmission in the frequency range that contributes most strongly to the emission.

Integrating a partial-wave spectrum over frequency gives  the power emitted in the related mode 
\begin{equation}
\label{eq}
\frac{dE}{dt}
\equiv
\int_0^\infty
\mathcal{P}{s\ell}(\omega)d\omega.
\end{equation}
The Hawking radiation power emitted by the black hole should be measured by summing over all $\ell$.  Physically, a stronger $dE/dt$ means a more efficient evaporation
process and hence a shorter lifetime because  the evaporation rate of the black hole is directly related to the energy emission rate \cite{Page:1976df}.

\begin{figure*}[htbp]
\centering
\includegraphics[height=4cm]{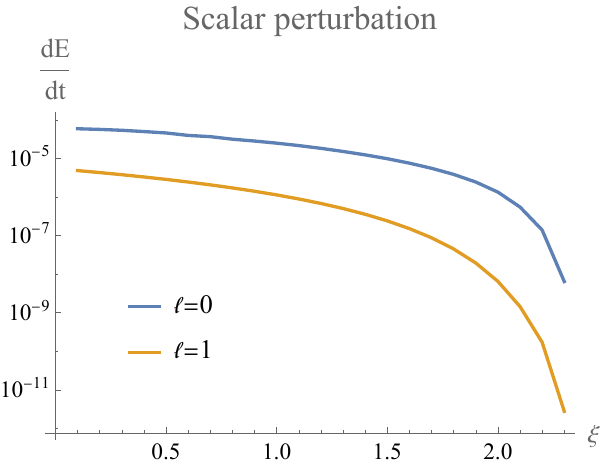}\hspace{0.1cm}
\includegraphics[height=4cm]{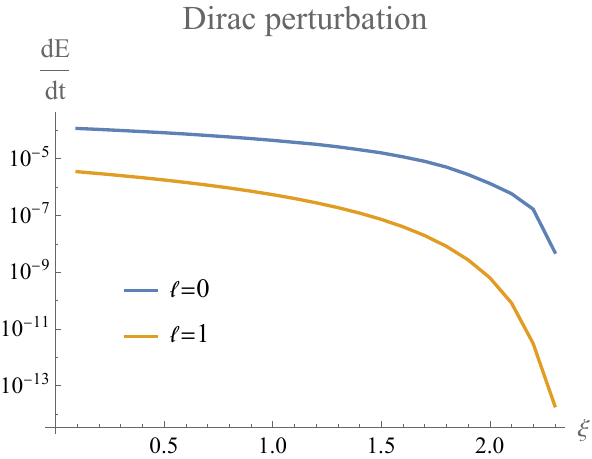}\hspace{0.1cm}
\includegraphics[height=4cm]{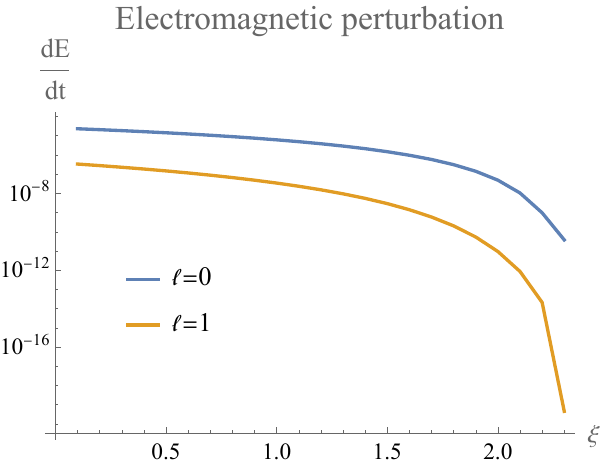}
\caption{\label{EER total}Total emission rate $dE/dt$ as a function of  $\xi$ for various fields with $s=0$ (left), $s= 1/2$ (middle) and $s= 1$ (right).}
\end{figure*}

To explicitly show the effect of $\ell$, we plot the estimated powers carried by the fist two individual angular modes for each field in FIG. \ref{EER total}. In all cases, the energy emission rate decreases with increasing $\xi$. For each field,  by comparing the curves with different angular modes, we find that the lowest radiative partial wave carries more power than the next one over the displayed parameter range. Therefore,  in terms of observation, this intuitively would mean that, such EYM black hole with a larger $\xi$ in the early universe could survive due to the low evaporation rate.  Especially, in the extremal case, the total EER  approaches zero as expected, meaning that such extremal black holes almost do not evaporate and thus they are often considered to live forever if they are in complete isolation, similar to the case in extremal RN black hole \cite{Adams:2000ax}.

It is noted that the lowest scalar contribution is particularly sensitive to the treatment of low frequency transmission. So the local features in the blue curve of the left plot in , saying the small fluctuation,  require independent numerical checks before they can be interpreted as physical effects. The overall suppression of the displayed emission channels is consistent with the decrease in the Hawking temperature, despite the weaker and more field dependent changes in the greybody factors.

\section{Conclusion and discussion}\label{sec:conclusion}

In this work, we investigated massless scalar, electromagnetic, and Dirac test fields around the non-minimal EYM black hole, focusing on their QNM frequencies, greybody factors and Hawking emission. The corresponding effective potentials exhibit a single barrier structure outside the event horizon. For the parameter range considered, the positive potentials and the damped QNMs support mode stability in these test field perturbations. The differences between the fields arise from their angular and spin dependent contributions to the effective potentials, which are particularly important at low multipoles.

We firstly examined the dependence of the QNM spectrum on the non-minimal coupling parameter $\xi$, the multipole number $\ell$, and the overtone number $n$. The lowest scalar mode is particularly sensitive to the coupling, consistent with the absence of a centrifugal contribution to its effective potential. The lower Dirac modes vary more weakly, but this weak dependence does not persist throughout the spectrum. Higher overtones generally exhibit stronger variations with $\xi$, including pronounced non-monotonic behavior in the scalar and Dirac sectors. The electromagnetic overtones also develop non-monotonic changes in their real part of frequencies, while their damping rates generally decrease as $\xi$ increases.

This study extended the comparison of perturbative responses beyond the spectra examined in Refs.~\cite{Jusufi:2020odz,Gogoi:2024qnm}. The enhanced sensitivity of higher overtones is qualitatively consistent with the scalar overtone outburst reported in Refs.~\cite{Lutfuoglu:2025ljm,Ji:2025qnm} and with earlier studies of near-horizon deformations \cite{Konoplya:2022hll,Konoplya:2022pbc}. In particular, the results contributed to compare this sensitivity across the scalar, electromagnetic, and Dirac sectors under the same background parameters, and they show that the response of the fundamental mode alone cannot characterize the effect of the non-minimal coupling on the full spectrum, and overtone modes show significant difference.

Furthermore, our results on the multipole dependence provides a complementary perspective. The real frequencies increase with $\ell$ for all three fields, whereas the damping rates approach their large-$\ell$ behavior differently. In the cases examined, the scalar damping rate decreases with increasing $\ell$, the electromagnetic damping rate increases, and the Dirac damping rate changes only weakly. At $\ell=100$, the frequencies agree with the leading eikonal description determined by the angular velocity and Lyapunov exponent of the unstable circular null orbit \cite{Cardoso:2008bp}. Increasing $\xi$ raises the orbital angular velocity and reduces the Lyapunov exponent, leading to higher oscillation frequencies and weaker damping in this regime. Thus, the field dependent behavior at low multipoles gives way to a common leading geometric description, with the remaining differences associated with subleading angular and spin-dependent terms.

For each fixed partial wave, the greybody factor increases with frequency and approaches unity in the high frequency limit. Increasing $\ell$ shifts the transmission transition toward higher frequencies, reflecting the stronger angular barrier. By comparison, we also found that the corrections induced by $\xi$ are weaker and depend on the field and multipole. Increasing $\xi$ suppresses transmission for the electromagnetic and Dirac modes considered, as well as for higher scalar multipoles. But the lowest scalar mode behaves differently, as its transmission probability can increase slightly over some frequency intervals. These changes depend on the full shape of the effective potential, including its height and width.
It is worthwhile to mention that our frequency dependent transmission estimates complement the greybody factor bounds obtained in Ref.~\cite{Al-Badawi:2023qnm}. A comparison with the scalar transmission enhancement reported in Ref.~\cite{Lutfuoglu:2025ljm} also requires attention to the quantities held fixed, in which that authors  varied the coupling at fixed horizon radius, whereas our calculations kept the mass and magnetic charge fixed. These choices describe different sequences of black hole geometries. A quantitative comparison therefore requires matching the dimensionless parameters and frequency normalization before interpreting differences in the coupling dependence.

Finally, we studied the Hawking emission spectra which reflect both the transmission probabilities and the thermal occupation factors. For the modes considered, the spectra rise from small values at low frequencies, reach a maximum, and then decay through thermal suppression. Increasing $\ell$ lowers the spectral peak and shifts it toward higher frequencies, showing that the suppression of transmission outweighs the increase in angular multiplicity in the parameter range examined.  Increasing $\xi$ reduces the frequency integrated partial wave powers for the electromagnetic and Dirac fields and for higher scalar multipoles. The lowest scalar mode exhibits a mild non-monotonic dependence, consistent with competing changes in its transmission probability and the thermal factor.

Several extensions would sharpen this comparison and help us further understand the current black hole as well as the related gravity theory. Firstly, direct integration of the scattering equations, supplemented by low-frequency asymptotics, would improve the determination of transmission probabilities in the regime where  WKB approximation becomes less reliable \cite{Konoplya:2019hlu}.  Secondly, a converged sum over the relevant partial waves and particle species would then be needed to determine the total luminosity. Assessing evaporation further requires an evolution of the mass and charge at fixed theory coupling. The last but not the least, an analysis of the coupled metric and YM perturbations derived from the underlying non-minimal action is needed to connect the test field results with the stability of the full system and its gravitational wave ringdown, which we hope to study elsewhere.

\begin{acknowledgments}
This work is partly supported by Natural Science Foundation of China under Grants No. 12375054 and the Postgraduate Research \& Practice Innovation Program of Jiangsu Province under Grant No. KYCX25\_3923.
\end{acknowledgments}

\bibliography{ref}
\bibliographystyle{utphys}

\end{document}